\pdfoutput=1

\documentclass[12pt, a4paper]{article}

\usepackage{notes2bib}
\usepackage{amsmath}
\usepackage{amsfonts}
\usepackage{amssymb}
\usepackage{bbm}
\usepackage{verbatim}
\usepackage{dsfont}
\usepackage{booktabs}
\usepackage{slashed}
\usepackage{bm}
\usepackage{enumitem}

\usepackage{epsfig}
\usepackage[table]{xcolor}
\usepackage{graphicx}
\usepackage[]{caption}
\usepackage[listofformat=empty,subrefformat=empty]{subfig}

\usepackage{float}
\usepackage{overpic}

\usepackage{enumerate}
\usepackage{hhline}
\usepackage{multirow}
\usepackage[hidelinks]{hyperref}

\usepackage{cite}
\usepackage{a4wide}
\usepackage{bbold}

\usepackage{geometry}
\renewcommand{\baselinestretch}{1.2}

\DeclareMathOperator{\tr}{\text{tr}}

\newcommand{\Mp}{M_{\mathrm{P}}}

\renewcommand{\tr}{\text{tr}}

\newcommand{\da}{\Delta_a}

\newcommand{\HH}{{\rm HH}}

\definecolor{darkgreen}{HTML}{3C8031}
\begin{document}

\thispagestyle{empty}

\begin{flushright}
DESY 26-041 
\end{flushright}
\vskip .8 cm
\begin{center}
  {\Large \bf Density matrix of de Sitter JT gravity}
    %as\\ a model for quantum cosmology}
  %\\[.2cm] 
 \\[12pt]

\bigskip 
{
{\bf{Wilfried Buchm\"uller}\footnote{E-mail:
    wilfried.buchmueller@desy.de}}
and
{\bf{Alexander Westphal}\footnote{E-mail:
    alexander.westphal@desy.de}}
}
\bigskip\\[0pt]
\vspace{0.23cm}
{\it Deutsches Elektronen-Synchrotron DESY, Notkestr. 85,
  22607 Hamburg, Germany}
\\[20pt] 
\bigskip
\end{center}

\date{\today}

\begin{abstract}
\noindent
Jackiw-Teitelboim (JT) gravity in two-dimensional de Sitter space 
is an intriguing toy model for a quantum mechanical description
of an inflationary phase of the universe, including initial
conditions. Starting from
exact solutions of the Wheeler-DeWitt equation, we study a
conditional density matrix of the system. We find that the ground
state is a mixed state, rather than a pure Hartle-Hawking state.
Our results are consistent with the semiclassical double-trumpet amplitude, and
with recent work on complex geometries containing bra-ket wormholes.
We also analyze semiclassical wave functions
for metric, dilaton, and an additional inflaton field.
The probability distribution
for the size of the universe is flat. 
\end{abstract}

\newpage 
\setcounter{page}{2}
\setcounter{footnote}{0}
{\renewcommand{\baselinestretch}{1}\tableofcontents}

\section{Introduction}
\label{sec:introduction}

The no-boundary proposal for a ``wave function of the universe''
\cite{Hartle:1983ai} has inspired four decades of quantum cosmology
(for reviews and references, see, for example,
\cite{Halliwell:1989myn,Lehners:2023yrj}). But the debate over the proposal
continues, and the goal to derive testable predictions, consistent
with the observed cosmic microwave background,  remains challenging
\cite{Maldacena:2024uhs}.

In recent years, important progress has been made in two-dimensional (2d)
cosmological toy models. A particularly interesting example 
is Jackiw-Teitelboim (JT) gravity \cite{Teitelboim:1983ux,Jackiw:1984je}. The
model is exactly solvable \cite{Henneaux:1985nw,Louis-Martinez:1993bge} and,
like all 2d dilaton-gravity theories, its minisuperspace version
already contains all physical information
\cite{Louis-Martinez:1993bge}. Over the past years, JT gravity has been
studied in detail in anti-de Sitter ($\text{AdS}_2$) space (for a
review, see, for example, \cite{Mertens:2022irh}),
and, more recently, the no-boundary wave function has also been
computed in de Sitter ($\text{dS})$ JT gravity at large field values 
\cite{Maldacena:2019cbz,Cotler:2019nbi}. This is achieved
by reducing the path integral for the wave function to
a path integral over the Schwarzian degrees of freedom
of a boundary curve, as in $\text{AdS}_2$ gravity
\cite{Stanford:2017thb}. Summing up an infinite series of 
extrinsic curvature terms of the boundary curve, the asymptotic
form of the wave function has been extended to the entire
field space \cite{Iliesiu:2020zld}. Correspondingly, solutions to the
Wheeler-DeWitt equation with Schwarzian asymptotic behaviour have been
analysed \cite{Buchmuller:2024ksd}.

As first noted in \cite{Iliesiu:2020zld}, in JT gravity the
Hartle-Hawking wave function is singular at the de Sitter radius.
This has been criticised in \cite{Fanaras:2021awm} since
it points toward a source not contained in the no-boundary proposal,
see also \cite{Buchmuller:2024ksd}. As a consequence, the
Hartle-Hawking wave function in JT gravity is not normalisable
\cite{Nanda:2023wne,Collier:2025lux,Dey:2025osp}).
It is therefore widely regarded as unphysical. The
singularity of the wave function is not visible in a semiclassical
saddle point approximation. Cosmological application of JT gravity
have been considered in extensions with conformal matter fields
\cite{Anninos:2024iwf}, and in Unimodular JT gravity \cite{Alexandre:2025rgx}.

In this paper we consider a version of JT gravity that corresponds to
`half reduction' obtained from three-dimensional de Sitter space, contrary to
the `full reduction' model derived from the metric of a
Schwarzschild-dS black hole (see, for example, \cite{Svesko:2022txo}).
Both, scale factor and dilaton take positive values only. In
reductions from Kantowski-Sachs cosmology \cite{Kantowski:1966te}
where the universe has the topology of $S^1\times S^2$, the dilaton
parametrises the size of the $S^2$ \cite{Fanaras:2021awm, Fanaras:2022twv}.
In general,
semiclassical wave functions depend on initial conditions and on
quantum corrections to the classical WKB wave functions. We analyse
the general structure of semiclassical wave functions by means of the
characteristics of the WDW equation for JT gravity and an extension
with an inflaton scalar field.

Similar to a wave function, also a density matrix can be defined as 
a gravitational path integral \cite{Page:1986vw, Hawking:1986vj}.
A no-boundary density matrix has been constructed for the observable
subregion of the universe \cite{Ivo:2024ill}. In general,
a density matrix receives contributions from disconnected as
well as connected geometries like bra-ket wormholes
\cite{Chen:2020tes}. Since an observer is confined to the universe, only
conditional probabilities are meaningful \cite{Page:1986vw}.

Recently, a conditional density matrix has been computed for
a complex geometry including bra-ket wormholes, and it was found
that the connected contribution dominates over the disconnected
contribution \cite{Fumagalli:2024msi}.
It turns out that, in the semiclassical approximation, this
conditional density matrix is consistent with the double-trumpet
amplitude computed in \cite{Cotler:2019nbi}, following earlier work
in Euclidean $\text{AdS}_2$ \cite{Saad:2019lba}. Disconnected and
connected contributions to the density matrix can also be constructed
starting from exact solutions of the WDW equation
\cite{Buchmuller:2024ksd}. In the semiclassical approximation, we
again find a result consistent with the bra-ket wormhole geometry.
As we shall see, this density matrix indeed satisfies the criterion for
mixed states\footnote{It has been argued that in JT gravity the path
integral with two closed boundaries factorises and that the Hilbert
space is one-dimensional \cite{Usatyuk:2024isz,Harlow:2025pvj}; for
recent related work in 4d de Sitter, see
\cite{Abdalla:2026mxn,Nomura:2026igt}. According to our analysis such
a factorisation does not occur.}. We therefore propose that the ground
state of de Sitter JT gravity is a mixed state described by a conditional density matrix rather than a pure
Hartle-Hawking state.

In JT gravity we are able to calculate this conditional density matrix of the universe in terms of an exact transition amplitude from the path integral. This amplitude involves a parameter $h_0$ corresponding to the initial size of the JT de Sitter universe. The choice of $h_0$ is related to the choice of a boundary condition imposed on a putative pure ground-state wave function~\cite{Buchmuller:2024ksd}. In our proposal of a conditional density matrix instead of a pure ground state, we view this dependence on the parameter $h_0$ as a quantum mechanical degeneracy. This implies that in constructing the mixed-state density matrix we trace over all possible boundary conditions labeled by $h_0$.

The resulting conditional mixed-state density matrix of the JT de Sitter universe has the interesting feature that  real-valued prefactors of the transition amplitudes cancel. For 4d de Sitter Hartle-Hawking pure ground states, such prefactors have been argued to produce exponentially strong biases toward small vacuum energy and against long-lasting slow-roll inflation (see, for example,~\cite{Maldacena:2024uhs} for a review, and~\cite{Abdalla:2026mxn} for a recent new treatment).

The paper is organised as follows.
We introduce and study our proposal of a conditional density matrix for
JT gravity and its properties in section~\ref{sec:dm}. The semiclassical
result for a bra-ket geometry is compared with a double-trumpet
amplitude and the density matrix obtained from exact solutions
of the WDW equation. Here the degeneracy of the ground state plays
a crucial role. In appendix~\ref{A:msDM} it is shown that the
semiclassical density matrix indeed satisfies the criterion for
a mixed state. In section~\ref{sec:semiJT} we construct the general semiclassical
wave function for JT gravity in terms of the integration constants of
the characteristics of the WDW equation. The general wave function
is compared with solutions of the WDW equation that contain quantum corrections.
These results are extended to JT gravity
with an additional inflaton field. Details of the construction are
given in appendices~\ref{A:PJTinf} and~\ref{A:dPsemiBranching}. Fluctuations of the inflaton field
and the suppression of the probability for large universes are
discussed in section~\ref{sec:fluc}. We conclude in section~\ref{sec:conclusion}.

\section{Wave function vs. density matrix}
\label{sec:dm}

\subsection{Hartle-Hawking wave function}

Jackiw-Teitelboim gravity \cite{Teitelboim:1983ux,Jackiw:1984je}
in de Sitter space is defined by the Lorentzian action
\begin{equation}\label{SJT}
  S_G[g,\phi] = \frac{1}{4\pi} \int_{\mathcal{M}}d^2x \sqrt{g} \phi (R - 2\lambda^2) +
  \frac{1}{2\pi}\int_{\partial\mathcal{M}} d\theta \sqrt{h} \phi K \ .
\end{equation}
Here $g$, $R$, $\lambda^2$, $h$ and $K$ denote  metric tensor, 
Ricci scalar and cosmological constant, and induced metric and extrinsic curvature on the boundary $\partial\mathcal{M}$, respectively. 
Compared to pure gravity with cosmological constant,
the action depends linearly on a dilaton field $\phi$.
In minisuperspace, which in 2d is just a gauge choice
\cite{Louis-Martinez:1993bge}, the metric with lapse function $N$, and $h=a^2$,
\begin{equation}\label{mLd2}
  ds^2 = -N^2(t)dt^2 + a^2(t) d\theta^2 \ , \quad 0\leq \theta < 2\pi \ ,
\end{equation}
yields the Lorentzian action 
\begin{equation}\label{ILjt}
  I_G[h,\phi] = \int dt N\left(-\frac{1}{N^2} \dot{a} \dot{\phi} - \lambda^2 a\phi
    \right) \ .
  \end{equation}
This implies the Hamiltonian constraint
    \begin{equation}\label{jtHC}
\dot{a} \dot{\phi} - \lambda^2 a\phi = 0 \ ,
\end{equation}
and the corresponding WDW equation ($N=1$)
\begin{equation}\label{wdwh}
  \left(\frac{\partial^2}{\partial h\partial\phi}
    +\frac{\lambda^2}{2} \phi\right)\Psi(h,\phi) = 0 \ .
\end{equation}

In de Sitter space the definition of a ground state, a state of
`minimal excitation', is a subtle question. A leading candidate is the
no-boundary proposal \cite{Hartle:1983ai} that defines the `wave
function of the universe' as a path integral
\begin{equation}
\Psi^\HH(h,\phi) =
\int^{(h,\phi)}[Dg][D\phi] \exp{(iS[g,\phi])}  \label{piHH}
\end{equation}
over complex manifolds $\mathcal{M}$ with metric $g$ and dilaton $\phi$ which match the 
values $(h,\phi)$ at a boundary $\partial\mathcal{M}$.

The path integral \eqref{piHH} has been computed in a saddle-point
approximation. The geometry corresponds to a half-hyperboloid of de
Sitter space matched to a half-sphere at the equator. Integrating the
classical field equation in the complex $t$-plane from the `south
pole' of the half-sphere to
the boundary in de Sitter space, and computing the quantum corrections
${\mathcal O}(\hbar)$, one obtains a semiclassical Hartle-Hawking wave function of 2d  de Sitter space in JT gravity
\cite{Buchmuller:2024ksd},
\begin{equation}\label{psisc}
\begin{split}
  \Psi^\HH_{sc}(h,\phi) &= C_{sc}(h,\phi) \exp{\left(-i\lambda\phi\sqrt{h-h_c}\right)} \ , \\
 C_{sc}(h,\phi) & = \frac{e^{\phi_0}}{\lambda\sqrt{h-h_c}}\ , \quad
  h > h_c = \lambda^{-2} \ .
\end{split}
\end{equation}
Here, $\phi_0$ is the value of the dilaton field
at the south pole of the half-sphere, and $\sqrt{h_c}$ is the radius
of the circle where sphere and hyperboloid match. At fixed dilaton,
$\phi=\phi_b$, and $h \gg h_c$, the square of the wave function
provides a probability distribution for universes of size $\sqrt{h}$.
Another solution of the WDW equation is the real wave function $\Psi + \Psi^*$,
analogous to the Hartle-Hawking wave function
\cite{Hartle:1983ai}.

The JT action is linear in the dilaton field $\phi$. This allows to
integrate out the bulk dilaton field, and to reduce the path integral
\eqref{piHH} to an integral over the Schwarzian degrees of freedom
of the boundary specified by the choice $\phi=\phi_b$. This yields
an exact result for the wave function at large field values
\cite{Maldacena:2019cbz,Cotler:2019nbi}.  One
finds a semiclassical wave function like Eq.~\eqref{psisc},
with $C_{sc}$ replaced by\footnote{The effect of the Schwarzian
  fluctuations on the prefactor depend on the relation between
  path-integral wave function and WDW wave function; here we use the original result in \cite{Maldacena:2019cbz}. The connection
proposed in \cite{Cotler:2024xzz}, which is adopted in appendix~G of
\cite{Maldacena:2019cbz} (v4), would modify the Hartle-Hawking wave
function in the semiclassical regime by a factor $\sim (\phi /(\lambda a))^{1/2}$, i.e., the wave function would
fall off faster. This would have no significant effect on the results
in this paper. Note that the $\phi$-dependence
  differs for different factor ordering in the WDW equation; for a
  discussion, see \cite{Buchmuller:2024ksd}.} $C_{sch}$
  \cite{Maldacena:2019cbz}
\begin{equation}\label{psisch}
\begin{split}
 \Psi^\HH_{sch}(h,\phi) &= C_{sch}(h,\phi) \exp{\left(-i\lambda\phi\sqrt{h-h_c}\right)} \ , \\
C_{sch}(h,\phi)  &=  
\frac{C_0}{\phi}\left(\frac{\phi}{\lambda\sqrt{h}}\right)^{3/2} \ .
\end{split}
\end{equation}

One can also find exact solutions of the WDW equation with Schwarzian
asymptotic behaviour \cite{Iliesiu:2020zld,Fanaras:2021awm}. They can
be expressed in terms of the transition amplitude\footnote{Here we have
  projected on one sign of the
  extrinsic curvature of the future surface; without projection one
  would obtain the real amplitude $\langle h,\phi|h_0,0\rangle_+ +
  \langle h_0,0|h,\phi\rangle_-$.}
\begin{equation}
\begin{split}
  \langle h,\phi|h_0,0\rangle_+ &=
  \int_{(h_0,0)}^{(h,\phi)}[Dg][D\phi']\exp{(iS_G[g,\phi'])} \\
  &= \langle h_0,0|h,\phi\rangle_-^* \ .
\end{split}
\end{equation}
In JT gravity, the integral is Gaussian, and one obtains
\cite{Halliwell:1990tu, Buchmuller:2024ksd,Honda:2024hdr}
\begin{equation}\label{ta+}
\langle h,\phi|h_0,0\rangle_+= H^{(2)}_0(\lambda \phi(h-h_0)^{1/2}) \Theta(h-h_0) \ .
\end{equation}
A wave function with Schwarzian asymptotic behaviour is obtained by
convoluting the transition amplitude with a singular boundary
condition at $\phi=0$,
\begin{equation}\label{psidh0}
  \Psi_+(h,\phi) = \int dh' \langle
  h,\phi|h',0\rangle_+\partial_{h'}\Psi(h',0) \ ,
  \quad \Psi(h',0) = -\delta(h'-h_0) \ .
\end{equation}
This yields the wave function
\cite{Buchmuller:2024ksd}
\begin{equation}
  \begin{split}
  \Psi_+(h,\phi;h_0) &= \partial_h H^{(2)}_0(\lambda\phi(h-h_0)^{1/2})
  \Theta(h-h_0) \\
  &= - \frac{\lambda\phi}{2(h-h_0)^{1/2}}
H^{(2)}_1(\lambda\phi(h-h_0)^{1/2})\Theta(h-h_0)\ ,
\end{split}
\end{equation}
where we have dropped a singular piece at $h=h_0$.\footnote{The real wave function reads
  $\Psi = \Psi_++\Psi_- = - \lambda\phi(h-h_0)^{-1/2}
  J_1(\lambda\phi(h-h_0)^{1/2})\Theta(h-h_0)$; note that the singular
  terms of $\Psi_+$ and $\Psi_-$ at $h=h_0$ cancel.
Recently, also the analogous wave function for
Euclidean $\text{AdS}_2$ has been derived \cite{Griguolo:2025kpi}.}
At large $h$, the Schwarzian scaling with $h$ is reproduced,
\begin{equation}\label{HHwf}
\Psi_+(h,\phi;h_0) \sim 
   \frac{1}{\phi}\left(\frac{\phi}{\lambda\sqrt{h}}\right)^{3/2}
   \exp{\left(-i\lambda\phi\sqrt{h}\left(1 -
         \frac{h_0}{2h}\right)\right)}\ .
 \end{equation}
 For $h_0=h_c$, this agrees with the large-$h$ Hartle-Hawking wavefunction in Eq.~\eqref{psisch} including the Schwarzian quantum corrections: $\Psi_+(h,\phi;h_c)\sim \Psi^\HH_{sch}(h,\phi)$.

 At $h\sim h_0$, the wave function has a pole,
 \begin{equation}
   \Psi_+(h,\phi;h_0) \sim \frac{1}{h-h_0} \ ,
 \end{equation}
 and it is a solution of an inhomogeneous WDW equation with a singular
 source at the boundary $\phi = 0$ placed at $h=h_0$. A possible physical realisation of such singular sources may arise from end-of-the-world branes; see, for example, the boundary proposal~\cite{Friedrich:2024aad}. A singularity of this type is expected
 since the WDW equation \eqref{wdwh} has a conserved Klein-Gordon
 current \cite{Buchmuller:2024ksd}.  By contrast, the no-boundary
 proposal requires regular solutions of a homogeneous WDW equation.
    Since the wave function $\Psi_+$ is not
 normalisable, it is widely regarded as unphysical (see, for example,
\cite{Fanaras:2021awm,Nanda:2023wne,Collier:2025lux}).

\begin{figure}[t]
 \centering
 \includegraphics[width = 0.385 \textwidth]{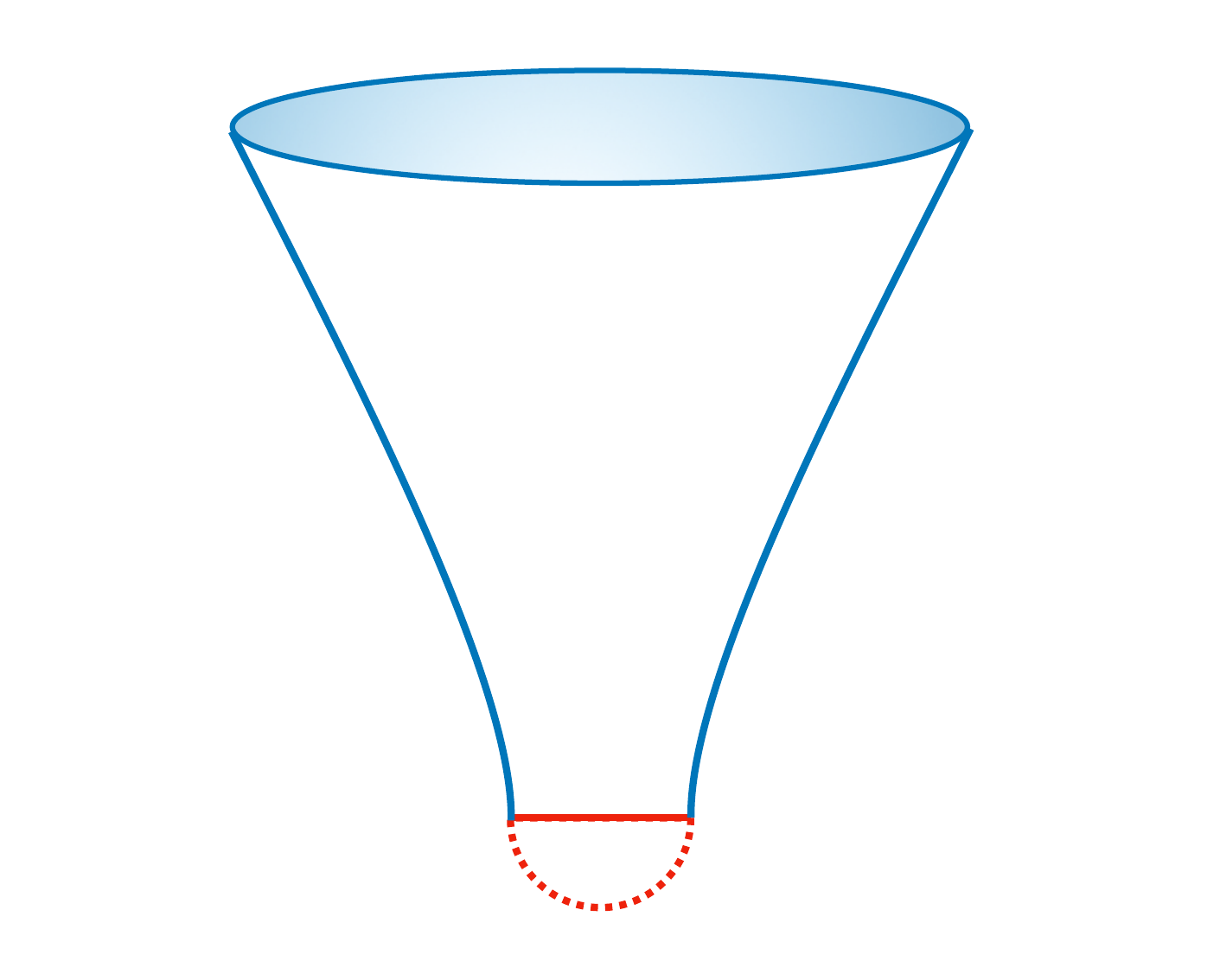} 
 \hspace{2cm}
 \includegraphics[width = 0.415 \textwidth]{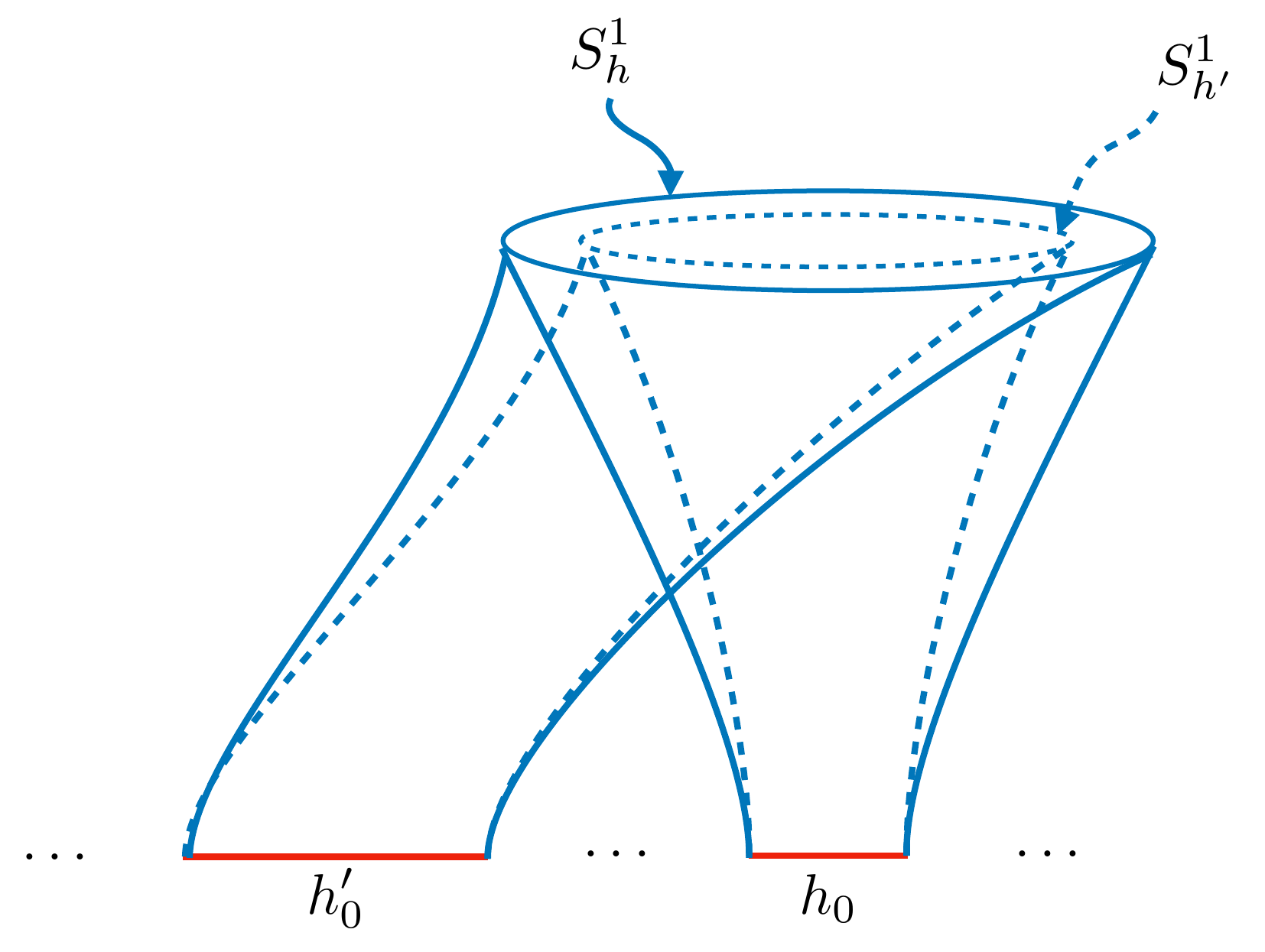} 
 \caption{Left: complex Lorentzian/Euclidean geometry underlying no-boundary wave function. Right: double-trumpet configuration for two different values of $h_0$.}
\label{fig:WF1}
\end{figure}

\subsection{JT density matrix}

A remarkable property of the wave function $\Psi_+(h,\phi;h_0)$ is the
dependence on the parameter $h_0 > 0$, a crucial difference compared
to the Hartle-Hawking wave function. There the parameter is fixed,
$h_0 = h_c = \lambda^{-2}$,
which is a consequence of the complex saddle-point
geometry (see Fig.~\ref{fig:WF1}). The appearance of the parameter $h_0$ is a direct
consequence of the invariance of the WDW equation \eqref{wdwh}
w.r.t. translations in $h$. Physically, $\sqrt{h_0}$ corresponds to
the `initial size' of the universe, with subsequent de Sitter
expansion. 

This invariance leads to a family of pure ground-state wave functions
labeled $h_0$. This is connected to the fact that picking a state with
definite $h_0$ corresponds to choosing a definite boundary condition
$\Psi(h',0)=-\delta(h'-h_0)$. However, we do not have a theory or
dynamical principle selecting a unique boundary condition. We
therefore propose to interpret this family of ground-state wave functions as a degeneracy, and to trace out all possible choices of boundary conditions, that is, all choices of $h_0$. This leaves us with a mixed-state density matrix of the universe, instead of a pure ground-state wave function.

A density matrix can be defined starting from the path integral for
the transition amplitude \cite{Page:1986vw,Hawking:1986vj}
\begin{equation}
  \langle h,\phi|h',\phi'\rangle_+ =
  \int_{(h',\phi')}^{(h,\phi)}[Dg][D\phi']\exp{(iS_G[g,\phi'])} \ .
\end{equation}
Unlike ordinary
quantum mechanical systems, the universe has no external observer.
Hence, only a conditional density matrix is a meaningful concept
where a transition amplitude is considered subject to a suitable
condition that selects a subspace of $[h,\phi]\cup[h',\phi']$
\cite{Page:1986vw}.

The above reasoning suggests to impose
the condition $\phi = \phi'=\phi_b$ on the two boundary states. To
incorporate the trace over the initial universe sizes $h_0$ we use
the quantum-mechanical superposition principle \cite{Hawking:1979ig} with
$\phi = 0$ as an intermediate surface and write the transition
amplitude as
\begin{equation}\label{TAI}
  \langle h,\phi_b|h',\phi_b \rangle = \int dh_0
  \langle h,\phi_b|h_0,0 \rangle \langle h_0,0|h',\phi_b \rangle \ .
\end{equation}
Projecting again on one sign of the extrinsic curvature at the surface
$(h,\phi_b)$, we arrive at a density matrix for expanding
universes\footnote{Note that $\rho_+$ corresponds to a density matrix of a
  quantum system where the different components have equal weight;
see, for example, \cite{Sakurai:2011zz}.}
\begin{equation}\label{rhodh0}
  \rho_+(h,\phi_b;h',\phi_b) = N_b^{-1}\int
  dh_0\rho^{(h_0,\phi_b)}_+(h,h') \ , 
\end{equation}
with
\begin{equation}
\begin{split}  
\rho^{(h_0,\phi_b)}_+(h,h') &=
\langle h,\phi_b|h_0,0\rangle_+\langle h_0,0|h',\phi_b \rangle_- \ ,\\
 N_b &= \int dh_0 \tr(\rho^{(h_0,\phi_b)}_+) \ , \\
\tr(\rho^{(h_0,\phi_b)}_+) &= \int dh |H^{(2)}_0(\lambda\phi_b\sqrt{h-h_0})|^2 \ .
\end{split}
 \end{equation}
 Note that the contribution from the region $h \sim h_0$ to the $h$-integral
 for $\tr(\rho^{(h_0,\phi_b)}_+)$ is finite because the transition
 amplitude has only a logarithmic singularity, contrary to the wave
 functions which have a pole at $h \sim h_0$. At large $h$, the
 integrand of $\tr(\rho^{(h_0,\phi_b)}_+)$ behaves as $1/\sqrt{h}$.
 The divergence at large $h$ will disappear in an extension of the model that
 includes reheating by its leading effect of ending the inflationary de Sitter phase.
 We shall represent this by introducing a
cutoff $\sqrt{h_{\text{max}}}=L_{\text{max}}$. In the integral
\eqref{rhodh0} the integration range for $h_0$ remains to be
specified. Without a complete theory for the density matrix, we shall
consider the classically forbidden domain $h_0 \in [0,h_c)$.

\begin{figure}[t]
 \centering
 \includegraphics[width = 0.4 \textwidth]{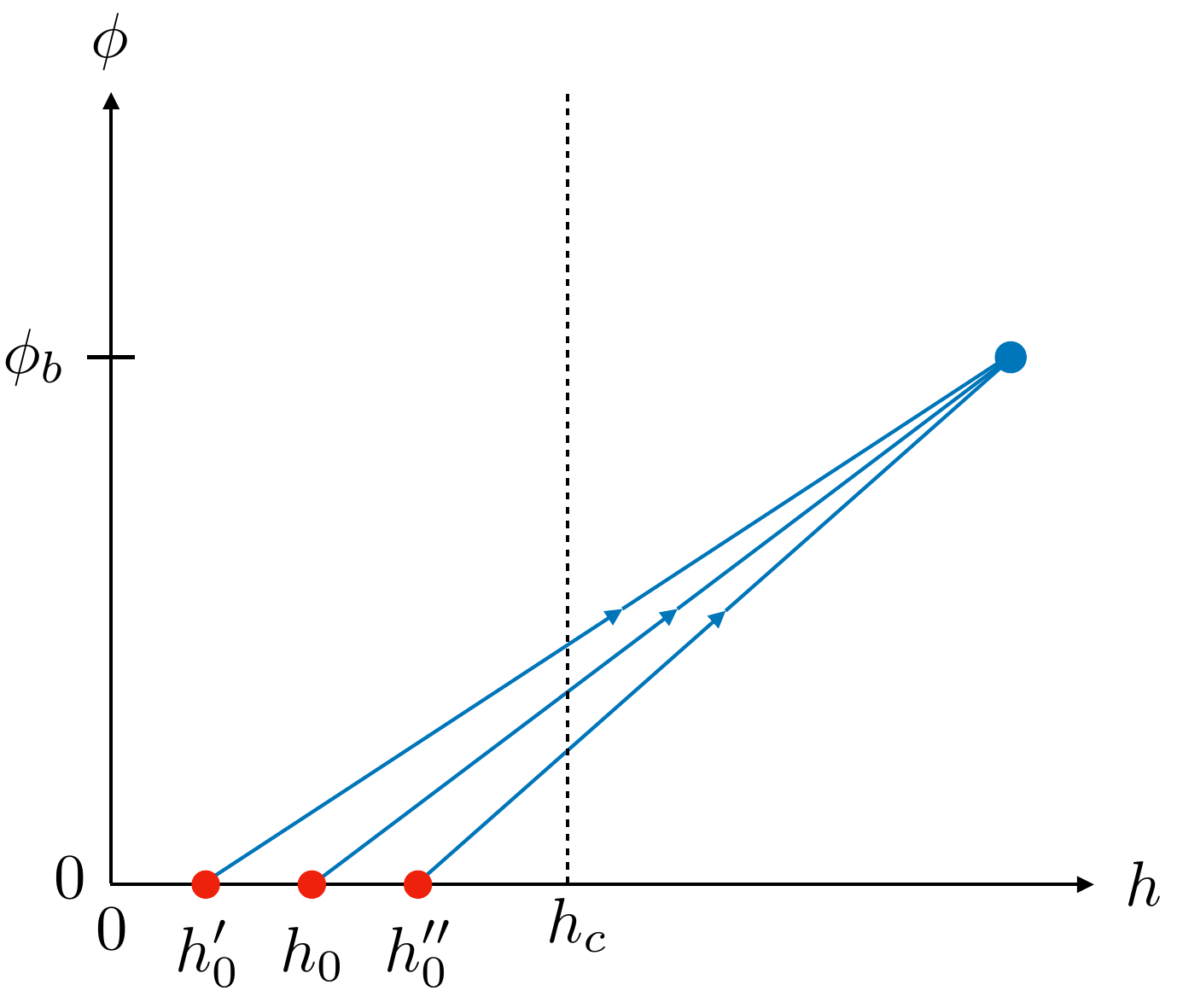} 
 \hspace{2cm}
 \includegraphics[width = 0.405 \textwidth]{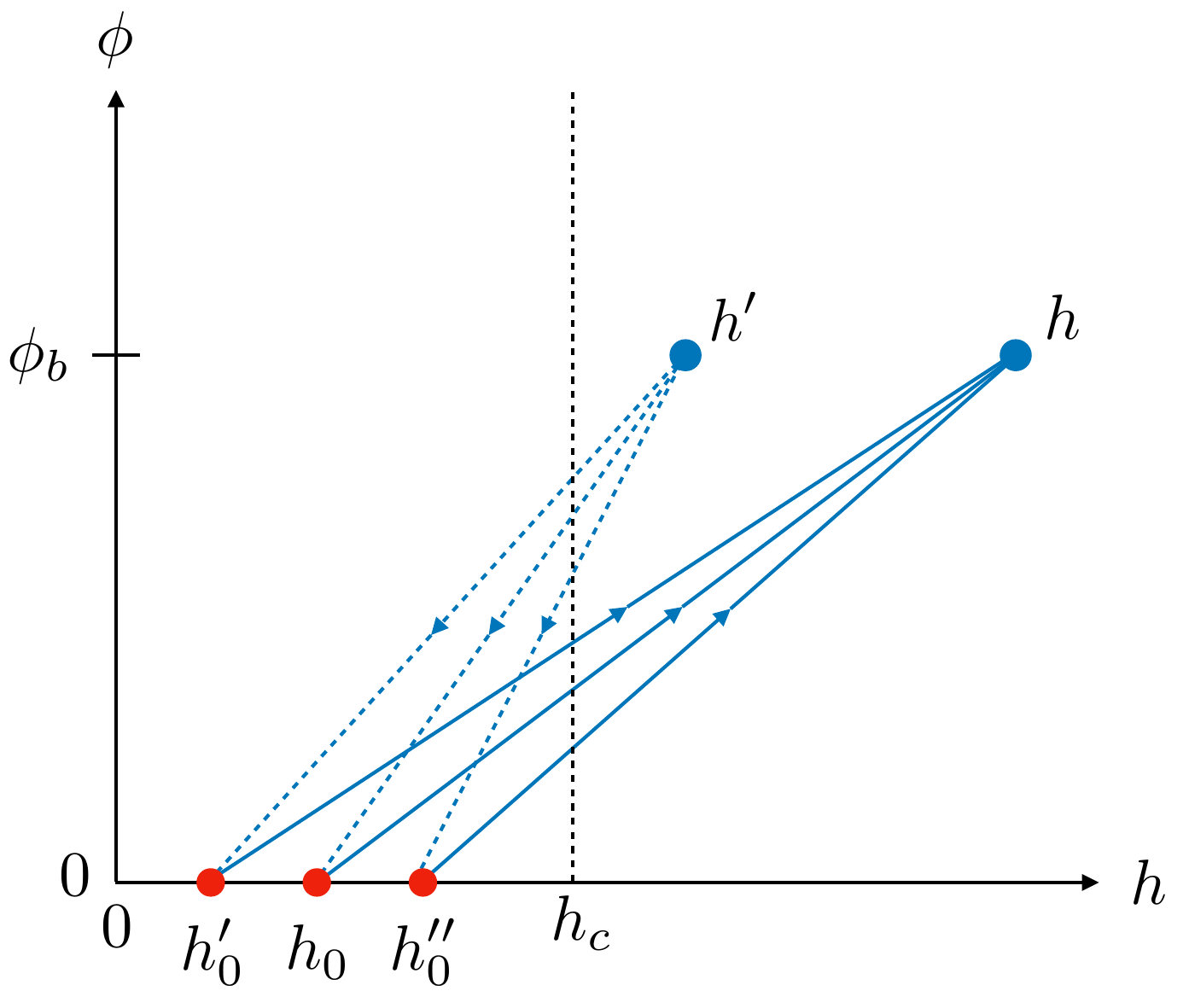} 
 \caption{Left: transition amplitudes for different values of $h_0$. Right: double-trumpet amplitudes for different values of $h_0$.}
\label{fig:WF2}
\end{figure}

In the semiclassical regime, $h \gg h_0$, the transition amplitude
\eqref{ta+} behaves as
\begin{equation}\label{FT}
\langle h,\phi|h_0,0\rangle \sim 
   \frac{1}{\phi}\left(\frac{\phi}{\lambda\sqrt{h}}\right)^{1/2}
   \exp{\left(-i\lambda\phi\sqrt{h}\left(1 -
         \frac{h_0}{2h}\right)\right)}  \ .
 \end{equation} 
This corresponds to the `future-trumpet' amplitude discussed in
 \cite{Cotler:2019nbi}, which is related by analytic continuation to
 the trumpet amplitude in Euclidean $\text{AdS}_2$
 \cite{Saad:2019lba}. It is less singular than the wave function at
 small $h$, which can be traced back to the fact that, compared
 to the wave function \eqref{HHwf},  one more
 Schwarzian fluctuation mode contributes.
 From the asymptotic form of the future-trumpet
amplitude \eqref{FT} one obtains for the density matrix in
the semiclassical regime ($h,h'\gg h_0$),
\begin{equation}\label{rho+}
  \begin{split}
  \rho^{(h_0,\phi_b)}_+(h,h') \sim 
   &\ \frac{1}{\lambda\phi_b^2}\left(\frac{\phi_b^2}{\sqrt{h}\sqrt{h'}}\right)^{1/2} \\
   & \ \times \exp\left(-i\lambda\phi_b\left(\sqrt{h} - \sqrt{h'} -
       \frac{h_0}{2}\left(\frac{1}{\sqrt{h}}-\frac{1}{\sqrt{h'}}\right)\right)\right)
               \ .
\end{split}
     \end{equation} 
The expression \eqref{rhodh0} with $\rho^{(h_0,\phi_b)}_+$ given
by Eq.~\eqref{rho+} yields a connected contribution to the density
matrix which does not factorise like a contribution from a pure state (see Fig.~\ref{fig:WF2}).
It satisfies the criteria for a density matrix (see, for example,
\cite{Landau:1991wop}): $\rho_+ = \rho_+^\dagger$,
$\tr(\rho_+) = 1$, and, as shown in appendix~\ref{A:msDM},
$\tr(\rho_+^2) < 1$. Hence, $\rho_+$ indeed
describes a mixed state.

The density matrix $\rho_+$ yields a probability
distribution for the size of the universe
in the semiclassical regime.
We integrate $h_0$ over the classically forbidden domain, $h_0 \in
[0,h_c)$, and regularise the divergent normalisation factor by a
cutoff $L_{\text{max}}$ (see appendix~\ref{A:msDM}). Such divergences at large scale factor for probability measures on de Sitter space, and relatedly, for slow-roll inflation describing quasi-dS, are well-known and form the basis of the so-called `measure problem' of eternal inflation whose discussion lies beyond the scope of this paper (for a review see, for example,~\cite{Freivogel:2011eg}). One then
obtains a scale-factor probability distribution $dP_\rho$ from the diagonal element of $\rho$ ($h=a^2$),
\begin{equation}\label{P+con}
\begin{split}
dP_{\rho,+}(a|\phi_b) &=\rho_+(h,\phi_b;h,\phi_b) dh\\
&= N_b^{-1}\int_0^{h_c} dh_0
\rho^{(h_0,\phi_b)}_+(h,h)\ dh\\
&\sim L^{-1}_{\text{max}}\  da\ .
\end{split}
\end{equation}
Note that the distribution is flat in $a$.

This can be compared with a conceivable contribution from the
Hartle-Hawking wave function conditioned on $\phi=\phi_b$ with the Schwarzian correction given by~\eqref{psisch}, which agrees with $\Psi_+(h,\phi;h_c)$ where $h_0=h_c$ is fixed. As this is a pure state, its conditional density matrix is given by
\begin{equation}\label{rhodhcHH}
  \rho^\HH_{sch}(h,\phi_b;h',\phi_b) = N_{\HH}^{-1}\Psi^\HH_{sch}(h,\phi) \Psi^{\HH\,*}_{sch}(h',\phi')\Big|_{\phi=\phi'=\phi_b} \ , 
\end{equation}
with $N_\HH=\int_{h_{\rm min}}^\infty dh|\Psi^\HH_{sch}(h,\phi_b)|^2$. Here the prefactor $C_{sch}(h,\phi)$ enforces a scaling $|\Psi^\HH_{sch}(h,\phi_b)|^2\sim |\Psi_+(h,\phi_b;h_c)|^2\sim C_0(h-h_c)^{-3/2}$. This renders the $h$-integral in the normalisation factor $N_{\HH}$ convergent at large $h$. Instead, it requires regularising the divergence of
the $h$-integral at $h \sim h_c$ by a cutoff
$\sqrt{h_{\text{min}}}=L_{\text{min}}$ implying that $N_{\HH}\sim 1/\sqrt{h_{\text{min}}}=1/L_{\text{min}}$. Hence, one finds for the
Hartle-Hawking wave function
($h\gg h_c$),
\begin{equation}\label{P+dis}
\begin{split}
  dP_{\rho,+}^{\text{dis}}(a|\phi_b)&\equiv  \rho^\HH_{sch}(h,\phi_b;h,\phi_b) dh\\
  &=\left.N^{-1}_{\HH} |\Psi^\HH_{sch}(h,\phi)|^2\right|_{\phi=\phi_b} dh\\
  &\sim N^{-1}_\HH |\Psi_+(h,\phi_b;h_c)|^2\ dh\\
  &\sim \frac{L_{\text{min}}}{a^2}\ da\ .
  \end{split}
 \end{equation}
This `disconnected distribution' is suppressed by a factor $a^2$
compared to the connected
contribution \eqref{P+con}. 

We note here a crucial feature of the conditional density matrix constructed in Eq.~\eqref{rhodh0} by projecting onto a slice labeled by $\phi=\phi_b$. Due to normalisation factor $N_b^{-1}$ inevitably present in this construction, the resulting conditional density matrix and the probability distribution in $a$ computed with it have no dependence on the real prefactor $C_0$ present in the Hartle-Hawking wave function $\Psi^\HH_{sch}$, as this cancels out within its associated pure-state conditional density matrix. This property is built into the structure of a conditional density matrix and will become crucial later in comparing scale-factor probability distributions constructed from a pure state conditioned onto $\phi=\phi_b$ using the measure $dP(a|\phi_b)=|\Psi(h,\phi)|^2\big|_{\phi=\phi_b} dh$  with those constructed from a conditional density matrix as given above.

\subsection{Density matrix from complex geometries}

Recently, a conditional density matrix has been evaluated for JT
gravity \cite{Fumagalli:2024msi}, based on a complex geometry with bra-ket wormholes
\cite{Chen:2020tes, Fumagalli:2024msi}. For global de Sitter and a
geodesic circle (the `bottle-neck') with circumference $2\pi \alpha$,
$\alpha \in [0,\infty)$, this density matrix reads in the semiclassical
regime $a,a' \gg \alpha$\cite{Fumagalli:2024msi},
\begin{equation}\label{rhoalpha}
  \rho^{(\alpha)} (a,\phi_b;a',\phi_b) =
  \frac{1}{2}\left(\frac{\phi_b^2}{aa'}\right)^{1/2}
  \exp{\Big(i\phi_b(a'-a) +i
      \alpha^2\frac{\phi_b}{2}\Big(\frac{1}{a'}
        - \frac{1}{a}\Big)\Big)} \ .
    \end{equation}
    The projection corresponds to fixing the value of the dilaton to
    $\phi_b$ on both boundary surfaces, and the integration over the
    modulus $\alpha$ is performed with the measure
    $\alpha d\alpha$ \cite{Saad:2019lba,Cotler:2019nbi},
    \begin{equation}
      \rho(a,\phi_b;a',\phi_b) = \int \alpha d\alpha
      \rho^{(\alpha)}(a,\phi_b;a',\phi_b)\ .
\end{equation}
It is remarkable that, up to a factor $\phi_b^{-2}$, the result \eqref{rhoalpha} is
identical\footnote{We believe that the sign of the second term in
  the exponential of Eq.~(5.2) in \cite{Fumagalli:2024msi}, and
  therefore in Eq.~\eqref{rhoalpha}, should be reversed.} to the expression
\eqref{rho+} derived in the previous section, with the identification
$\alpha = \sqrt{h}$, which parametrises the radius of the boundary
circle in the two approaches. 
In \cite{Fumagalli:2024msi}, one considers a Wigner distribution, a
Fourier transform
of the semiclassical density matrix \eqref{rhoalpha} w.r.t. $a_-=a-a'$.
The double integral over $a_-$ and $\alpha$ is then evaluated at a
saddle point $\alpha(a_+,p,\phi_b)$, where $a_+ = a + a'$, and $p$ is the momentum
conjugate to $a_-$. The result is a classical probability distribution
on the phase space variables $a_+$ and $p$. Comparing this connected 
contribution to the disconnected one given by the product of
Hartle-Hawking wave functions, it is found that the probability distribution
is dominated by the connected contribution. This is consistent with
the result obtained in the previous section.

Up to a kinematical exponential factor, the expressions \eqref{rho+}
and \eqref{rhoalpha} also agree with the semiclassical global $\text{dS}_2$
 double-trumpet amplitude \cite{Cotler:2019nbi}. 
 In \cite{Cotler:2024xzz}, the double-trumpet is
 considered at infinity, $a\phi_b \rightarrow \infty$,
 with $\phi_b/a \equiv \Phi/(2\pi)^2$ fixed. The integral over
 $\alpha$ is then taken from $0$ to $\infty$. The resulting
 amplitude has a singularity at $a = a'$, which is treated by means of an
 $i\epsilon$-prescription\footnote{Note that the so-defined amplitude
 differs from the standard transition amplitude $\langle
 a,\phi_b|a',-\phi_b \rangle$.}.

\section{Semiclassical de Sitter JT wave functions}
\label{sec:semiJT}

Cosmological applications of JT gravity require fields in addition to
metric tensor and dilaton. For these more
complicated theories no exact solutions of the WDW equation are
available and, as a first step, one has to rely on semiclassical
approximations. In the following we therefore discuss a method
to determine the general semiclassical wave function based on
the characteristics of the partial differential equation for the
prefactor of the WKB wave function. We first explain the method
for JT gravity and then apply it to JT gravity with an inflaton.

\subsection{De Sitter JT gravity}
\label{sec:JT}

In minisuperspace the Lorentzian action~\eqref{ILjt} with lapse function $N$ and $a=\sqrt{h}$,
\begin{equation*}
  I_G[a,\phi] = \int dt N\left(-\frac{1}{N^2} \dot{a} \dot{\phi} - \lambda^2 a\phi
    \right) \ ,
  \end{equation*}
 yields the Hamiltonian constraint~\eqref{jtHC},
    \begin{equation*}
\dot{a} \dot{\phi} - \lambda^2 a\phi = 0 \ ,
\end{equation*}
and therefore  the WDW equation ($N=1$)
\begin{equation}\label{cfo}
  \left(\hbar^2\partial_a\partial_\phi
    +\lambda^2 a \phi\right)\Psi(a,\phi) = 0 \ ,
\end{equation}
where we have kept Planck's constant $\hbar$.

The solution to the equation of motion for the scale factor, $a(t) =
a_0\cosh{(\lambda t)}$, interpolates between
a circle of minimal radius $a_0$ at $t = 0$ and a circle of radius
$a > a_0$ at $t_a = \lambda^{-1}
\operatorname{arcosh}{(a a_0^{-1})}$ of the de Sitter hyperboloid. The corresponding solution for the dilaton field, satisfying
the constraint \eqref{jtHC} and the boundary condition
$\dot{\phi}(0) = \lambda\phi_0$, reads
\begin{equation}\label{phi0}
  \phi(t) = \phi_0 \sinh(\lambda t)\ .
%  \quad \phi_0 = \frac{\phi}{\sinh(\lambda t_a)} \ .
\end{equation}
Note that the trajectory in the $a$$-$$\phi$ plane
is determined by the integration constants $a_0$ and $\phi_0$,
\begin{equation}\label{aphi}
  \phi = \phi_0 \da a_0^{-1} \ , \quad \text{with} \quad
 \da = (a^2-a_0^2)^{1/2} \ .
\end{equation}
A boundary circle can be specified by fixing the variable $a$ or the variable $\phi$.
From Eqs.~\eqref{ILjt} and \eqref{jtHC} one
obtains the on-shell action \cite{Buchmuller:2024ksd}
\begin{equation}\label{osG}
  I_G^{os}(a,\phi) = -\lambda\phi\da \ , 
\end{equation}
which can also be directly read off from Eq.~\eqref{SJT} by using $R = 2\lambda^2$
and inserting the extrinsic curvature $K = -\lambda \da a^{-1}$.

In the semiclassical regime, for large values of $a$ and $\phi$, the
system is described by the WKB wave function\footnote{For a review and references, see, for example \cite{Halliwell:1992cj,Lehners:2023yrj}.}
\begin{equation}\label{psiJT0}
  \Psi_0(a,\phi) =
  \exp{\left(\frac{i}{\hbar}I_G^{os}(a,\phi)\right)} \ ,
\end{equation}
which solves the WDW equation to leading order $\hbar^0$. The
$\cal{O}(\hbar)$ correction yields a slowly varying prefactor $C(a,\phi)$,
\begin{equation}
\Psi(a,\phi) = C(a,\phi) \Psi_0(a,\phi) \ ,
\end{equation}
satisfying the linear partial differential equation (PDE)
\begin{equation}\label{pdeC1}
\left(\partial_\phi I_G^{os}\partial_a + \partial_a
  I_G^{os}\partial_\phi\right) z = -\partial_a\partial_\phi I_G^{os}
\ , \quad z = \ln{C} \ .
\end{equation}
Inserting the on-shell action \eqref{osG} yields 
\begin{equation}\label{pdeC2}
  \left(\Delta_a \partial_a + \phi a \Delta_a^{-1}
    \partial_\phi \right) z = -  a  \Delta_a^{-1}\ .
\end{equation}

The general solution to this PDE can be
found by determining the characteristics
$a(s)$, $\phi(s)$ and $z(s)$ that satisfy the ordinary differential
equations (see, for example, \cite{Courant:1965st})
\begin{equation}
\frac{da}{ds} = \Delta_a \ ,\quad
\frac{d\phi}{ds} = \phi a\Delta_a^{-1}\ ,\quad
\frac{dz}{ds} = - a \Delta_a^{-1}\ .
\end{equation}
The solutions are given by
\begin{equation}
  \begin{split}
&a = a_0\cosh{s} \ , \quad
\phi = \phi_0 \sinh{s} \ ,\\
&z = - \ln\sinh{s} + z_0 \ ,
\end{split}
\end{equation}
where $a_0$, $\phi_0$ and $z_0$ are integration constants.
The solutions $a(s)$ and $\phi(s)$ are identical
with the solutions to the equations of motion discussed above, so that
the parameter $\lambda^{-1}s$ can be identified with the coordinate time $t$. One
can also invert the above relations and express the
integration constants as functions of the field variables.
Since $d\phi_0/ds = 0$,
\begin{equation}\label{phi0}
  \phi_0(a,\phi) = \frac{\phi a_0}{\da}
\end{equation}
is a solution of the homogeneous PDE
\eqref{pdeC2}. $dz_0/ds = 0$ implies that $z(a,z_0)$
is a solution of the inhomogeneous PDE \eqref{pdeC2}.
Choosing $z_0 = \ln(f(\phi_0))$,  where $f$ is
an arbitrary function, one obtains for $C(a,\phi) = \exp{(z(a,\phi))}$,
\begin{equation}\label{CJT}
  C(a,\phi) = f(\phi_0(a,\phi))\frac{a_0}{\da}
= f\left(\frac{\phi a_0}{\da}\right) \frac{a_0}{\da} \ .
\end{equation}
One easily verifies that
$C(a,\phi)$ indeed satisfies the PDE~\eqref{pdeC2}.

To determine $C(a,\phi)$ in JT gravity, initial conditions and quantum
corrections have to be taken into account. The `no-boundary contour'
in the complex-time plane, which yields the Hartle-Hawking wave function
in the case of 4d de Sitter space, corresponds to $a_0 = \lambda^{-1}$ and
$f = \exp(\phi_0)$ in the case of JT
gravity~\cite{Buchmuller:2024ksd}.
On the contrary, the connected part of the complex bra-ket geometry
\cite{Fumagalli:2024msi} as well as exact solutions of the WDW equation
allow for arbitrary values of $a_0$.
In the Hartle-Hawking case, incorporating
the quantum fluctuations of the Schwarzian degrees of freedom on the boundary
\cite{Maldacena:2019cbz},  the comparison with the general form
\eqref{CJT} yields the prefactor
\begin{equation}\label{Cmty}
  C(a,\phi) = \frac{C_0}{\phi}\left(\frac{\phi a_0}{\da}\right)^{3/2} \ ,
 \end{equation}
or equivalently,
\begin{equation}\label{fmty}
  f(\phi_0) = C_0 \phi_0^{1/2}  = C_0 \left(\frac{\phi a_0}{\da}\right)^{1/2}\ .
\end{equation}
This result is also consistent with the semiclassical limit of an exact solution
of the WDW equation \cite{Buchmuller:2024ksd}.

The square of the wave function, $|\Psi|^2 = C^2$, provides a
measure on the congruence of classical trajectories
\cite{Halliwell:1992cj}. Using Eq.~\eqref{Cmty},
the Hartle-Hawking measure, and fixing $\phi=\phi_b$,
one obtains the probability distribution for the scale factor ($a \gg a_0$)
\cite{Buchmuller:2024ksd},
\begin{equation}\label{dP}
  dP(a|\phi_b)
  = \left.|\Psi|^2\right|_{\phi=\phi_b} dh= C_0^2\phi_b\left(\frac{a_0}{\da}\right)^3dh  
  \sim C_0^2\frac{da}{a^2} \ .
  \end{equation}
  The same distribution in $a$
  is obtained for the Klein-Gordon measure,
  however with a different dependence on $\phi_b$.
$dP(a|\phi_b)$ is interpreted as
the probability for finding a universe with a 1-geometry which is a
circle of size $a$ in the interval $(a,a+da)$ and with a given value
$\phi_b$ of the dilaton \cite{Maldacena:2019cbz}.

The prefactor also depends on the factor ordering, contrary to the
 oscillating exponential WKB factor. Changing from `canonical
 factor ordering' in Eq.~\eqref{cfo} to `Henneaux factor ordering',
 the WDW equation becomes\footnote{For a recent discussion of factor
   ordering in JT gravity, see \cite{Franken:2025unn}.}
\begin{equation}\label{Hfo}
  \left(\hbar^2 a \partial_a a^{-1} \partial_\phi
    +\lambda^2 a \phi\right)\tilde{\Psi}(a,\phi) = 0 \ .
\end{equation}
The partial differential equation \eqref{pdeC1} is then replaced by
\begin{equation}\label{pdeH}
\left(\partial_\phi I_G^{os}\partial_a + \partial_a
  I_G^{os}\partial_\phi\right) \ln{\tilde C}
= -\partial_a\partial_\phi I_G^{os}  + a^{-1} \partial_\phi I_G^{os} \ ,
\end{equation}
for which one finds the general solution
\begin{equation}
\tilde{C}(a,\phi) = C_0\frac{a}{\phi}\left(\frac{\phi a_0}{\da}\right)^{3/2}\, .
\end{equation}
This is also consistent with the semiclassical limit of an exact solution
of the WDW equation \cite{Buchmuller:2024ksd}. The result can be
directly obtained from Eq.~\eqref{Cmty} by using the relation between the wave
functions for the different factor orderings: $\tilde{\Psi} = a \Psi$
\cite{Iliesiu:2020zld}. Hence, for large $a$ the probability distribution is changed
by a factor $a^2$. The factor ordering also modifies the 
conserved Klein-Gordon current, however in such a way
that the probability distribution
remains unchanged \cite{Buchmuller:2024ksd}.

\subsection{Semiclassical wave functions with inflaton}
\label{sec:JTinf}

 We now extend JT gravity by adding an inflaton with linear
 potential\footnote{This is well motivated by various models of
   inflation.}, following \cite{Fumagalli:2024msi}. The corresponding action reads
 \begin{equation}\label{LM}
    I_M[a,\chi] = \int dt a \left(\frac{1}{2} \dot\chi^2 +
      \lambda^2(\kappa\chi - c) \right) \ ,
    \quad \kappa > 0 \ ,
 \end{equation}
 where we have introduced a linear inflaton potential with negative
 slope $\lambda^2\kappa$ and
 a `cosmological constant' $\lambda^2c$ corresponding to the
 potential at $\chi = 0$. Since the potential is unbounded from below,
 we consider $I_M(a,\chi)$ as an effective action for a dilaton with
appropriately chosen finite field range.   
 The equations of motion for the scale factor $a$, for inflaton and
 dilaton, and the Hamiltonian constraint are obtained from $I = I_G + I_M$,
\begin{align}
   &\ddot a - \lambda^2 a = 0 \ , \label{ema}\\
 &\ddot\chi + \frac{\dot a}{a} \chi - \lambda^2\kappa  = 0 \ , \label{emchi} \\ 
   &\ddot\phi -\lambda^2\phi +\frac{1}{2}\dot\chi^2
     + \lambda^2(\kappa \chi - c)  = 0 \  , \label{emphi} \\
   &\dot a \dot\phi -\lambda^2 a\phi -\frac{1}{2} a\dot\chi^2
     + \lambda^2 a(\kappa\chi - c) = 0 \ . \label{hcd}
 \end{align}
Note that these equation are not independent. For example,
Eq.~\eqref{emphi} follows from Eqs.~\eqref{ema}, \eqref{emchi} and \eqref{hcd}.

With $a = a_0 \cosh{\lambda t}$, the solution for inflaton
and dilaton read 
\begin{align}
 \chi &=  \chi_0 X(a) + \kappa \ln{(a a_0^{-1})}
        \ , \label{inf} \\
 \phi &= \Delta_a a_0^{-1} \Big(\phi_0-\frac12\left(\kappa^2+\chi_0^2\right)X(a)\Big) +\kappa^2 \ln{(a a_0^{-1})}+\kappa\chi_0 X(a)-c-\frac{\chi_0^2}{2} \ . \label{phi}
\end{align}
Here $\chi_0$ and $\phi_0$ are integration constants, and
we have defined\footnote{A further integration constant for $\chi(t)$ 
has been chosen such that $\chi(0) = 0$.} 
\begin{equation}
  X(a) =  \arccos{(a_0a^{-1})}\ .
\end{equation} 
Furthermore, we chose as the initial condition for $\dot\phi$: $\dot\phi\big|_{a=a_0}=\lambda(\phi_0+\kappa\chi_0)$. The solution~\eqref{phi} then satisfies both Eqs.~\eqref{emphi} and \eqref{hcd}.
From Eq.~\eqref{inf} one obtains for $a \gg a_0$, 
\begin{equation}\label{dclass}
  a \simeq a_0 \exp{\left(\kappa^{-1} \chi\right)} \ ,
\end{equation}
which means that the value of the inflaton field counts the number of
e-folds during the de Sitter expansion.

As in the case of pure JT gravity one can determine
the on-shell action $I^{os}$ by using the equations of motion.
This yields the result
\begin{equation}\label{osGM}
    I^{os}(a,\phi,\chi)  = I^{os}_G(a,\phi) + I^{os}_M(a,\chi)  \ ,
\end{equation}
with 
\begin{equation}\label{osM}
  (\lambda a_0)^{-1}I^{os}_M = - \left(c - \kappa\chi
    + \frac{\kappa^2}{2}\right)\da a_0^{-1}  + \frac{\kappa^2}{2} X
  + \frac{1}{2}\left(\chi -\kappa\ln{(a a_0^{-1})}\right)^2 X^{-1} \ ,
  \end{equation}
and $I_G^{os}=-\lambda\phi\da$, see Eq.~\eqref{osG}.

  The hamiltonian constraint \eqref{hcd} yields the WDW equation
\begin{equation}\label{cfod}
\left(\hbar^2\left(\partial_a\partial_\phi -\frac{1}{2a}\partial_\chi^2\right)
  +\lambda^2 a (\phi   - \kappa \chi + c) \right)\Psi(a,\phi) = 0 \ ,
\end{equation}
which is solved by the WKB wave function
\begin{equation}\label{psiaphichi}
  \Psi_0(a,\phi,\chi) = \exp{\left(\frac{i}{\hbar}I^{os}(a,\phi,\chi)\right)}
\end{equation}
to leading order $\hbar^0$. The semiclassical wave function is 
\begin{equation}
\Psi(a,\phi,\chi) = C(a,\phi,\chi) \Psi_0(a,\phi,\chi) \ ,
\end{equation}
where to $\cal{O}(\hbar)$ the prefactor satisfies the partial differential equation
\begin{equation}\label{pdedC}
\left(\partial_\phi I^{os}\partial_a + \partial_a
  I^{os}\partial_\phi-\frac{1}{a} \partial_\chi
  I^{os}\partial_\chi\right) z = -\partial_a\partial_\phi
I^{os} + \frac{1}{a}\partial_\chi^2 I^{os}\ , \quad z = \ln{C} \ .
\end{equation}
Inserting the on-shell action \eqref{osGM} yields 
\begin{equation}\label{pdgd}
 \begin{split}
 \Big(& \da \partial_a 
+ \Big(\Big(\phi + c - \kappa \chi
             + \frac{\kappa^2}{2}\Big) a\da^{-1}
          - \frac{\kappa^2}{2} a_0^2(a\da)^{-1}  \\
 &          + \kappa a^{-1}(\chi - \kappa \ln{(a a_0^{-1})})X^{-1}a_0
 + \frac{1}{2} (\chi - \kappa \ln{(a a_0^{-1})})^2
 X^{-2}a_0^2(a\da)^{-1} \Big)\partial_\phi\\
 &+ a^{-1} (\kappa \da +  (\chi  - \kappa \ln{(a a_0^{-1})})
 X^{-1}a_0)\partial_\chi\Big) z(a,\phi,\chi)\\
&        = - a \da^{-1} - a_0 a^{-1} X^{-1}\ .
\end{split}
  \end{equation}
  The general solution can again be obtained by using the method of 
  characteristics. There are now three integration constants,
  $\chi_0(a,\chi)$, $\phi_0(a,\phi,\chi)$ and
  $z_0(a,z)$. Eqs.~\eqref{inf} and \eqref{phi} yield $\chi_0$ and
  $\phi_0$. For $z_0$ one finds (see appendix~\ref{A:msDM}),
  \begin{equation}
    z_0 = z + \ln{(X\da a_0^{-1})} \ .
\end{equation}
 Choosing $z_0= F(\phi_0,\chi_0)$, where $F$ is an arbitrary function,
 yields the solution to~\eqref{pdgd}
  \begin{equation}
    C(a,\phi,\chi) = F(\phi_0  (a,\phi, \chi), \chi_0 (a,\phi,\chi)) a_0\da^{-1} X^{-1} \ .
  \end{equation}

For JT gravity with inflaton no exact solutions of the WDW equation are
known, even in the semiclassical regime. As long as the backreaction
of the dilaton on the metric is small, a reasonable ansatz is the
solution \eqref{fmty} for pure JT gravity, with $\phi_0(a,\phi)$ in Eq.~\eqref{CJT} 
replaced by $\phi_0(a,\phi,\chi)$. This yields
\begin{equation}\label{fmtyd}
  F(\phi_0 (a,\phi, \chi), \chi_0 (a,\phi,\chi)) \simeq f(\phi_0
  (a,\phi, \chi)) \ ,
\end{equation}
and therefore,
\begin{eqnarray}\label{Cmtyinf}
 C(a,\phi,\chi) &\simeq&
 f(\phi_0(a,\phi,\chi)) \frac{a_0}{\da X}\nonumber\\
 &=& C_0 \phi_0^{1/2} \frac{a_0}{\da X}\ .
 \end{eqnarray}
From Eqs.~\eqref{inf} and \eqref{phi} one obtains for the integration
constants $\chi_0$ and $\phi_0\,$,
\begin{align}
  \chi_0 (a,\chi) &= X^{-1} \left(\chi - \kappa\ln{(a}{a^{-1}_0)}
  \right)  \ , \ X(a) = - \frac{\pi}{2}+\mathcal{O}(a^{-1})  \\
  \phi_0 (a, \phi, \chi) &= \frac{\phi a_0}{\da}
   - \frac{\pi\kappa^2}{4}  - \frac{1}{\pi}(\chi - \kappa\ln{(a a_0^{-1})})^2 + \mathcal{O}(a^{-1}) \label{phi0inf}\ .
\end{align}
Here one had to be careful in picking the correctly-sided limit of the $\arccos$-function in $X(a)$, for details see appendix~\ref{A:dPsemiBranching}.

Using again the Hartle-Hawking measure, we obtain from
Eqs.~\eqref{Cmtyinf} and \eqref{phi0inf} for the
probability distribution, 
up to terms of relative order $\mathcal{O}(a^{-1})$,
\begin{align}\label{dPinfSC}
  dP(a,\chi|\phi_b) &=|C|^2 dh \nonumber\\
  &= C_0^2\phi_0 \frac{a_0^2}{\Delta_a^2}X^{-2}\nonumber\\
&\simeq \frac{4C_0^2}{\pi^2}  \left(\frac{\phi_b a_0}{\da}
- \frac{\pi\kappa^2}{4}  -
\frac{\kappa^2}{\pi}\left(\frac{\chi}{\kappa}
  - \ln{(a a^{-1}_0)}\right)^2\right) \frac{a^2_0}{\da^2}ada d\chi \ .
\end{align}
For $\kappa = 0$ one recovers the distribution \eqref{dP} of pure JT
gravity, up to a factor $4/\pi^2 = X^{-2}(0)$. An interesting
feature of the distribution is the local maximum in $\chi$ at
$\chi = \kappa\ln{(a a_0^{-1})}$, which corresponds precisely to the
classical solution \eqref{dclass}. At this maximum, the effect of the
inflaton on the distribution is a constant term $\propto \kappa^2$ in
\eqref{dPinfSC}: For $\phi_ba_0/\da > \kappa^2$ one obtains the fall-off
in $a$ of JT gravity, $dP \sim a^{-2}da$, whereas for large enough scale factor we reach $\phi_ba_0/\da < \kappa^2$ and $dP$ would turn negative. We  view this as a sign that the inflaton can no longer be treated as a mere perturbation. As long as the correction from the inflaton remains small, the above probability distribution reproduces the earlier results from pure JT gravity.

Moreover, for large deviations of
$\chi$ from the classical maximum the distribution \eqref{dPinfSC} is
unbounded from below and cannot be trusted. In this case the backreaction of the inflaton on the metric has to be taken into account in a full quantum mechanical calculation, which will change the
distribution \eqref{dPinfSC}. The distribution in the inflaton, i.e.
the number of e-folds, is flat.

\section{Fluctuations of the inflaton field}
\label{sec:fluc}

\subsection{WDW equation with an inhomogeneous inflaton}

The complete system of metric, dilaton and inflaton is described by
a functional WDW equation.
After a Fourier decomposition of the inflaton field\footnote{For global slicing one
  has $\chi(x) = \sum_k \chi_k\exp(ikx)$; for flat slicing the sum is
  replaced by $\int dk/(2\pi)$.}, canonical quantisation yields
the partial differential equation for the wave function ($\hbar = 1$),
\begin{equation}
  \left(\partial_a\partial_\phi + \frac{1}{2a}\sum_k\left(-\partial^2_{\chi_k}
  +k^2 \chi_k^2\right) +\lambda^2 a \phi   + \kappa a\chi_0 + a c
\right)\Psi[a,\phi;\{\chi_k\}] = 0 \ .
\end{equation}
As long as the backreaction on the metric is neglected, 
the inflaton is a free massless field in de Sitter space, and the
solution of the WDW equation factorises into the semiclassical wave
function of JT gravity and a product of wave functions for the momentum modes
of the inflaton,
\begin{equation}
  \Psi[a,\phi;\{\chi_k\}] = \Psi_0(a,\phi) \prod_k \Psi_k(a,\chi_k) \ .
\end{equation}
Here $\Psi_0$ is the WKB wave function \eqref{psiJT0} of JT gravity, and 
the wave functions $\Psi_k$ satisfy the Schroedinger equation for a
harmonic oscillator with frequency $|k|$,
\begin{equation}\label{HOinf}
i\partial_\eta \Psi_k = \frac{1}{2}\left(-\partial^2_{\chi_k} +
  k^2\chi^2_k\right)\Psi_k \ .
\end{equation}
Here $d\eta = a^{-1}(\lambda^2 a^2 - 1)^{-1/2}da$, which implies
\begin{equation}
\eta = - \operatorname{arc}\cos(\lambda a) \ , \quad \lambda a = \sin^{-1}(-\eta) \ .
\end{equation}
Eq.~\eqref{HOinf} is solved by (see, for example, \cite{LaFlamme:1990kd})
\begin{equation}\label{solC}
  \Psi_k(a(\eta),\chi_k) = N_kC_k^{-1/2}(\eta)\exp\left(\frac{i}{2}
    \frac{\dot{C_k}(\eta)}{C_k(\eta)} \chi_k^2\right) \ ,
\end{equation}
where $C_k$ satisfies the wave equation
\begin{equation}
  \ddot{C_k} + k^2 C_k = 0 \ .
\end{equation}
For $k>0$, the plane wave $C_k = \exp{(ik\eta)}$
yields a normalisable solution,
\begin{equation}
 \langle \Psi_k|\Psi_k\rangle = \int_{-\infty}^{\infty}d\chi_k |\Psi_k|^2 
= \int_{-\infty}^{\infty}d\chi_k |N_k|^2\exp(-k\chi_k^2) =1 \ .
 \end{equation} 
 Omitting half of the modes, $k < 0$, corresponds to the choice of
 a Bunch-Davies vacuum \cite{Birrell:1982ix}. An initial condition for the wave function
 $\Psi_k(a(\eta)),\chi_k)$ has to be specified at $\lambda a = 1$, i.e., at
 $\eta = -\pi/2$. Choosing $N_k = (k/\pi)^{1/4}\exp(-i\pi/4)$, one
 obtains as initial condition the familiar ground state wave
 function of a harmonic oscillator with frequency $\sqrt{k}$,
 \begin{equation}
   \Psi_k|_{\lambda a =1} = \left(\frac{k}{\pi}\right)^{1/4}
   \exp\left(-\frac{1}{2}k\chi_k^2\right) \ .
 \end{equation}
 At finite scale factor $a$, the wave function reads
 \begin{equation}\label{wfHO}
\Psi_k(a,\chi_k) = \left(\frac{k}{\pi}\right)^{1/4}
\exp\left(-\frac{i}{2}\left(\eta(a) + \frac{\pi}{2}\right)
-\frac{1}{2}k\chi_k^2\right) \ .
 \end{equation}
 
 The zero-mode $\Psi_0$ is not normalisable. However, starting from
 the action \eqref{LM} one  can construct the WKB wave function, 
 \begin{equation}
   \Psi_0(a,\chi) = \exp\left(\frac{i}{\hbar}I_M^{os}(a,\chi)\right) \ ,
   \end{equation}
where the on-shell action $I_M^{os}$ is given by Eq.~\eqref{osM}.
   Combined with the wave function \eqref{psiJT0}, one obtains
   the WKB wave function \eqref{psiaphichi} for scale factor, dilaton
   and inflaton, for which the structure
   of the quantum corrections has been analysed in
   section~\ref{sec:JTinf}.
   Note that the form of the wave function \eqref{solC} can also be
   used to study ground state and exited states of massive fields \cite{Birrell:1982ix}.
   
   An important quantity is the zero-point fluctuations of the
   inflaton field. Using the translation invariance of the expectation
   value, one obtains
   \begin{equation}
     \begin{split}
     \Delta^2_\chi(a) &= \langle \Psi|\Psi \rangle^{-1} \langle
     \Psi||\chi(\eta(a),x)|^2|\Psi \rangle \\
     &= \int \sum_k d\chi_k \Psi^*(a,\chi_k) \chi_k^2 \Psi(a,\chi_k) \ .
\end{split}
   \end{equation}
   Inserting the wave function \eqref{wfHO}, and replacing the sum 
   over momenta by an integral, which is a good approximation at large $a$, one
   finds the dimensionless power spectrum
   \begin{equation}
   \Delta^2_\chi(a) \equiv  \int d\ln k\ \Delta^2_\chi(a,k) = \int
   \frac{dk}{2\pi}\frac{1}{2k} \ .
 \end{equation}
 Hence, in two dimensions the power spectrum $\Delta^2_{\chi}(a,k)$ is constant.
 In particular, it is independent of the scale factor.

 \subsection{Probability distribution with inflaton}

In section~\ref{sec:JT} we have seen how the form of wave functions in
the semiclassical regime depends on the integration constants of the
characteristics of the WDW equation. This allowed us to estimate the
effect of an inflaton on the wave function as long as the backreaction
on the space-time geometry is small - we simply replaced the
integration constant without inflaton by the integration constant with
inflaton. The probability distribution $dP_+$ is proportional to the
integration constant $\phi_0(a,\phi_b) = \phi_b a_0/a$, with $a = \sqrt{h} \gg
a_0$ and $a_0 = \lambda^{-1}$. This suggests to estimate the inflaton
effect on the connected probability distribution by substituting again
$\phi_0(a,\phi_b)$ by $\phi_0(a,\phi_b,\chi)$ given in Eq.~\eqref{phi0inf}. 
This leads to the probability distribution ($a \gg a_0$)
\begin{equation}\label{dPinfcon}
  dP_{\rho,+}(a,\chi|\phi_b) 
\sim \frac{1}{\lambda^2\phi_b^2} \left(\frac{\phi_b}{\lambda a}
- \frac{\pi\kappa^2}{4}  -
\frac{\kappa^2}{\pi}\left(\frac{\chi}{\kappa}
  - \ln{(\lambda a)}\right)^2\right) dh d\chi \,\,,\,\, dh=2a da\,  \ .
\end{equation}
At the local maximum in $\chi$, and as long as the inflaton correction remains small, one obtains a flat distribution in $a$,
\begin{equation}\label{dPinfmax}
  dP_{\rho,+}(a,\chi_{max}|\phi_b) 
  \sim \frac{\phi_b/\lambda}{\lambda^2\phi_b^2} da\sim dP_{\rho,+}(a|\phi_b)  \ 
\end{equation}
matching the probability distribution Eq.~\eqref{P+con} from the connected piece of the conditional density matrix in pure JT gravity.

What is the origin of the difference between the probability distributions $dP_{\rho,+}$
\eqref{P+con} and $dP_{\rho,+}^{\text{dis}}$ \eqref{P+dis} arising from 
  the connected and the disconnected contribution to the density matrix,
  respectively? The square of the wave function $|\Psi|^2$ at large
  scale factor is determined by the fluctuations of the boundary curve,
  $dP_{\rho,+}^{\text{dis}} \sim a^{-2}da$. However, this Schwarzian boundary
  condition leads to a strong singularity at the de Sitter radius,
  which makes the interpretation of the Hartle-Hawking wave function
  in JT gravity very problematic \cite{Fanaras:2021awm}. 

 In the connected contribution with bra-ket wormholes the Lorentzian
 part of the complex geometry does not start at a fixed scale factor, but can take any value along the positive real
 axis; we made the choice $\sqrt{h_0}<\sqrt{h_c}=\lambda^{-1}$. The same is true for the double-trumpet amplitude
 \cite{Cotler:2019nbi}. The behaviour at large scale factor is again
 determined by the Schwarzian boundary fluctuations but for future
   and past trumpet amplitudes one more fluctuating mode  contributes. This
 changes the asymptotic behaviour of the amplitudes from $a^{-3/2}$ to
 $a^{-1/2}$ \cite{Saad:2019lba,Cotler:2019nbi}, which leads to
 the flat probability distribution $dP_{\rho,+}$.
 
 The exact solutions of the WDW equation have an integration
 constant $h_0 > 0$ that corresponds to the smallest size of the
 Lorentzian de Sitter hyperboloid. We interpret $h_0$ as a label of degenerate
 `ground-state wave functions' $\psi_{h_0}(h,\phi)=\langle
 h,\phi|h_0,0\rangle$. This suggests to build a density matrix
 by integrating products of ground state wave functions over $h_0$.
 Choosing the constant measure $dh_0$ leads to a density matrix
 in agreement with the double-trumpet amplitude and bra-ket wormholes.
 \\

\subsection{Exponential suppression of large universes}

Consider now a 4d de Sitter phase during slow-roll inflation, following
\cite{Maldacena:2024uhs}. It starts
at an inflaton value $\chi_*$, and an initial universe size
$a(\chi_*)$, when a characteristic momentum $k_*$ exits the horizon:
$k_*/a(\chi_*) = H(\chi_*)$, where $H$ is the Hubble parameter.
Suppose that after $N$ e-folds of expansion, the universe has reached the 
reheating surface with size $a(\chi_b) = H(\chi_*)^{-1}\exp(N)$.
In slow-roll inflation the number of e-folds is
determined by the inflaton values at beginning and end of inflation,
\begin{equation}
  N \simeq - \int_{\chi_*}^{\chi_b} d\chi \left(\partial_\chi \ln V\right)^{-1} \ ,
\end{equation}
where $V$ is the inflaton potential. Fixing the reheating surface at $\chi_b$,
the no-boundary wave function predicts the
probability for a universe with $N$ e-folds \cite{Hawking:1984hk} ($\Mp^2=8\pi$),
\begin{equation}
  |\Psi_N|^2 \sim \exp\left(\frac{24 \pi^2}{V(\chi_*)}\right) \ .
\end{equation}
A change in $N$ corresponds to a change of the initial inflaton
value. The relative probability can be written as \cite{Maldacena:2024uhs}
\begin{equation}\label{DeltaN}
  \frac{|\Psi_{N+\Delta N}|^2}{|\Psi_N|^2} \sim
  \exp\left(-\frac{2}{A_s}\Delta N\right) \ .
  \end{equation}
Here $A_s$ is the amplitude which, together with the spectral index
$n_s$, yields the curvature power spectrum that determines the
fluctuations in the cosmic microwave background (see, for example,
\cite{Baumann:2018muz}),
\begin{equation}
  \Delta^2_{\mathcal{R}} \simeq \left(\frac{H}{\dot\chi}\right)^2
  \Delta^2_\chi \simeq A_s \left(\frac{k}{k_*}\right)^{n_s-1} \ , \quad
\Delta^2_\chi = \langle \chi^2 \rangle = \left(\frac{H}{2\pi}\right)^2 \ ;
\end{equation}
$\Delta^2_\chi$ denotes the inflaton zero-point fluctuations.
Note, that   $\Delta^2_{\mathcal{R}}$ is approximately constant for
superhorizon scales, whereas  $\Delta^2_\chi$ does depend on the
scale factor. From Eq.~\eqref{DeltaN} one concludes that the small
value of $A_s$, measured by the CMB, implies a large exponential
suppression for the probability of large universes.

In two dimensions the situation is different. There are no curvature
perturbations, and even if one would weakly couple another field to
the inflaton, which could measure the zero-point fluctuations of the
inflaton at the reheating surface, this would provide no information
about the size of the universe since the fluctuations $\Delta^2_\chi$
do not depend on the scale factor. In 2d, the probability for a
universe with size $a$ at the reheating surface $\phi_b$ is given by
the value of the dilaton on the south pole of the Euclidean
half-sphere \cite{Buchmuller:2024ksd},
\begin{equation}
  |\Psi|^2 \sim \exp\left(2\phi_0(a,\phi_b)\right) \ .
\end{equation}
Using Eq.~\eqref{phi0}, and starting inflation at $H = \lambda$,
one obtains for large $N = \ln(\lambda a)$,
\begin{equation}
   |\Psi_N|^2 \sim \exp\left(2\phi_b\exp(-N)\right) \ .
\end{equation}
Hence, also in 2d large universes are exponentially suppressed,
\begin{equation}\label{eq:NsensJT}
  \frac{|\Psi_{N+\Delta N}|^2}{|\Psi_N|^2} \sim \exp\left(-2\phi_be^{-N}\Delta N\right)= \exp\left(-2\phi_0\Delta N\right)\ .
\end{equation}
This result does not change if instead of Eq.~\eqref{phi0} we use
$\phi_0(a,\phi,\chi)$ in Eq.~\eqref{phi0inf}, which includes the effect of the inflaton.

We now observe an analogy between the e-fold dependence of the probability distributions from 4d and 2d JT semiclassical Hartle-Hawking wave functions. Both distributions show exponential sensitivity to the total duration of inflation -- through the slow-roll based e-fold dependence of the $\exp(24\pi^2/V)$ prefactor of the Hartle-Hawking wave function in 4d, and through the e-fold dependence of the $\exp(\phi_0(a,\phi_b))$ prefactor of the Hartle-Hawking wave function in 2d JT gravity.

In 4d the size of the coefficient $1/A_s$ controlling the exponential e-fold dependence is directly related to the total duration (in e-folds) $N$ of inflation after the point of comparison (between inflationary histories lasting $N$ and $N+\Delta N$ e-folds). Taking the inflaton scalar potential $\frac12 m^2\chi^2$ as an example, we get  
\begin{equation}
1/A_s\sim10^{9} (60/N)^2\quad .
\end{equation} 
Hence, long-lasting inflationary histories have a probability distribution nearly flat in $N$ while inflation with short duration produces an exponentially strong bias towards fewer e-folds.

In 2d, the duration of inflation is fixed in terms of the dilaton evolution and is not tied to a slow-roll inflaton or its curvature perturbations (as these are absent here). Instead, the $\exp(\phi_0(a,\phi_b))$ prefactor generates an exponential e-fold dependence whose coefficient by Eq.~\eqref{eq:NsensJT} is controlled by the initial dilaton value $\phi_0$ which in turn is a combination $\phi_b\exp(-N)$ of the measured value $\phi_b$ of the dilaton and the total duration of inflation (which is exact dS expansion in 2d) $N$. Thus, here as well we see that long-lasting inflation has a probability distribution nearly flat in $N$ while inflation with short duration produces a strong exponential pressure towards fewer e-folds.

\section{Summary and conclusions}
\label{sec:conclusion}

The definition of a ground state, a state of `minimal excitation' , is
a subtle problem of gravity in de Sitter space.  Since forty years,
the no-boundary proposal of Hartle and Hawking is a leading candidate
although a number of issues remain to be settled. These include
problems of the path integral for complex manifolds, the validity of
saddle-point approximations, and, on the phenomenological side,
the realisation of a sufficiently long period of inflation (for a
discussion and references, see \cite{Maldacena:2024uhs}).

In recent years new insights have been gained from studying nearly de
Sitter space in two-dimensional Jackiw-Teitelboim gravity. In this
model the asymptotic behaviour of the no-boundary wave function for
large field values can be computed exactly
\cite{Maldacena:2019cbz,Cotler:2019nbi}, which leads to a prediction
for the probability distribution of the size of the universe.
However, the wave function has a power-singularity at the de Sitter radius
\cite{Iliesiu:2020zld}. Hence, it is not a solution of the WDW equation and not
normalisable \cite{Fanaras:2021awm}. 

The starting point of this paper are the exact solutions of the WDW
equation with Schwarzian asymptotic behaviour that were analysed in
\cite{Buchmuller:2024ksd}. Their characteristic feature is the
dependence on a parameter $\sqrt{h_0}$ that corresponds to the minimal size of
the Lorentzian hyperboloid. By contrast, for the no-boundary wave
function this parameter is fixed to the de Sitter radius $\sqrt{h_c} =
\lambda^{-1}$. One can now consider superpositions of wave functions
with varying $h_0$. Real wave functions, i.e., superpositions of
outgoing and incoming branches like the original Hartle-Hawking wave
function, have no singularity. It is not clear, however, how to select
from the many possible linear combinations a ground state. Moreover,
one has to worry about the needed projection to outgoing or incoming
branches. This may be realised by decoherence\cite{Halliwell:1989vw},
but will again require a source, contrary to solutions of the
WDW equation.

In this paper we interpret the dependence of the WDW solutions
on the initial size $\sqrt{h_0}$ of the de Sitter hyperboloid as a
degeneracy. Motivated by this, we propose a mixed state as the ground state. This is obtained
by i) tracing over $h_0$ in the density matrix and ii) conditioning onto a value of the dilaton $\phi=\phi_b$. For each $h_0$, the corresponding contribution to the resulting conditional density matrix consists of a coupled
outgoing and an incoming branch, similar to a double-trumpet
amplitude. As a consequence, the Schwarzian fluctuations lead to a fall-off
of the density matrix at large scale factors less strongly
than the square of the no-boundary wave function. Correspondingly, the
singularity at the de Sitter radius is only logarithmic and therefore
integrable. Our results are consistent with previous calculations for
complex geometries, the semiclassical double-trumpet amplitude
\cite{Cotler:2019nbi}, and the semiclassical density matrix obtained
for a Hartle-Hawking geometry with bra-ket wormholes
\cite{Fumagalli:2024msi}. In our approach, the weighting of the
 contributions to the density matrix has to be specified.
For the complex geometries, the weighting is fixed and corresponds to
the simplest possibility: tracing out $h_0\,$.
In appendix~\ref{A:msDM} we have shown that this definition indeed leads
to a mixed state.
From the diagonal element of the density matrix one obtains a flat
probability distribution for the scale factor of the universe, $dP_{\rho,+}(a|\phi_b)
\sim da$. This has to be compared with the probability distribution
obtained from a pure-state density matrix of the no-boundary wave function, $dP_{\rho,+}^{\rm dis}(a|\phi_b) \sim a^{-2} da\;$.

For most dilaton-gravity theories in $\text{dS}_2$ with additional
fields no exact solutions of the WDW equation are available. Here,
semiclassical methods are still useful. We have discussed
a general method to obtain semiclassical wave functions, which is
based on the characteristics of the WDW equation. We have used this
method to construct semiclassical wave functions for JT gravity with an
inflaton field. They are obtained in terms of the integration constants
of the characteristics, as explained in appendix~\ref{A:PJTinf}. The
results can be used to obtain approximate probability distributions
w.r.t. scale factor and dilaton. The limited domain of validity of
these distributions shows where the method breaks down.

Finally, we note a crucial difference between a scale-factor probability distribution computed from a pure ground-state wave function $\Psi$, conditioned onto a slice in field space $\phi=\phi_b\,$,
\begin{equation}
dP(a|\phi_b)=|\Psi(h,\phi)|^2\Big|_{\phi=\phi_b} dh\, ,
\end{equation}
and the distribution computed in terms of the associated conditional density matrix. This density matrix reads
\begin{equation}\label{eq:dPrho}
dP_\rho(a|\phi_b)=\rho(h,\phi_b;h,\phi_b) dh \equiv \frac{\Psi(h,\phi) \Psi^*(h,\phi')\Big|_{\phi=\phi'=\phi_b}}{\int dh  \Psi(h,\phi)\Psi^*(h,\phi')\Big|_{\phi=\phi'=\phi_b}} dh\quad .
\end{equation}
We denote by $\phi$ the observable onto whose measurement,
$\phi=\phi_b\,$, both $\Psi$ and $\rho$ are conditioned.

In our example of JT gravity, this observable $\phi$ is the dilaton. Consider a ground state wave function given by the Hartle-Hawking state $\Psi^\HH_{sch}(h,\phi)$ and the pure-state density matrix built from it. We can now condition both onto $\phi=\phi_b$. This produces a conditional pure-state density matrix of the Hartle-Hawking wave function. Compare now $dP$ with the $dP_\rho$ from the conditional pure-state density matrix. We then see that 
\begin{equation}
dP(a|\phi_b)\sim C_0^2 \frac{da}{a^2}\, .
\end{equation}
Conversely, this dependence on the real prefactor $C_0$ of $\Psi^\HH_{sch}(h,\phi)$ cancels out in $dP_\rho\,$.  In the semiclassical Hartle-Hawking wave function for JT gravity in de Sitter space, we approximate this prefactor by the $\exp(\phi_0)\,$. Hence, we see that the exponential dependence of this real prefactor on the total duration of inflation (the e-fold number $N$) cancels in a probability distribution $dP_\rho(a|\phi_b)$ built from the conditional density matrix. This cancellation is absent in a probability distribution built using the measure $dP(a|\phi_b)$ for a pure ground state wave function.

We now consider possible implications of this observation for the
wave function of 4d de Sitter space. First, we note that the above
benefits only accrue once we construct a conditional density matrix of
the universe by projecting onto a slice in field space.\footnote{We leave the effects of 
    an observer for
  obtaining a consistent quantum description of de Sitter
  space~\cite{Blommaert:2025bgd,Blommaert:2026lvp} as a task for
  future work.} Unlike 2d de Sitter space described by JT gravity, 4d
de Sitter space in pure Einstein gravity has no dilaton-like
slice-labeling degree of freedom. Therefore, applying our JT derived
reasoning to 4d requires replacing pure 4d de Sitter with a quasi-dS
space-time described by a slow-rolling scalar inflaton field
$\chi$. The values of the inflaton now provide the
slice-labeling onto which we can condition a density matrix.
The real $\exp(24\pi^2/V(\chi))$ prefactor of the 4d 
  Hartle-Hawking wave function plays no role if the connected part
  of the density matrix dominates. It also cancels in the conditional
  density matrix for the pure Hartle-Hawking wave function due to the structure of Eq.~\eqref{eq:dPrho}.
Therefore, unlike the exponential bias in favour of short inflation
present in 4d in the probability distribution $dP(a|\chi_b)$, the
scale-factor probability distribution
$dP_\rho(a|\chi_b)$ built from a conditional density matrix
is intrinsically free of this bias.

%As a final takeaway, we may wonder if the full notion of the `no boundary proposal' were to be such that it does not only select saddle points of the path integral describing smooth `cap-off' manifolds with no boundary to the Euclidean past, but that it should be understood as demanding in addition a `no boundary condition' rule: that one shalt not specify a definite boundary condition in terms of a pure ground state wave function, but instead just a statistical ensemble of possible boundary conditions in terms of a conditional mixed-state density matrix.

\subsection*{Acknowledgments}
We thank Arthur Hebecker for collaboration in the initial phase of the
project and for comments on the manuscript, and Jean-Luc Lehners,
Juan Maldacena and Guilherme Pimentel for valuable discussions. AW is partially supported by the Deutsche Forschungsgemeinschaft under Germany’s Excellence Strategy - EXC 2121 “Quantum Universe” - 390833306, by the Deutsche Forschungsgemeinschaft through a German-Israeli Project Cooperation (DIP) grant “Holography and the Swampland”, and by the Deutsche Forschungsgemeinschaft through the Collaborative Research Center SFB1624  “Higher Structures, Moduli Spaces, and Integrability’'.

\appendix

\section{Mixed state density matrix}
\label{A:msDM}

In this section we discuss the semiclassical part of the density
matrix.

For $h \gg h_0$, the asymptotic behaviour of the transition amplitude
is given by
\begin{equation}\label{ampsemi}
\langle h,\phi_b|h_0,0\rangle \sim 
   \left(\frac{2}{\pi\lambda\phi_b\sqrt{h}}\right)^{1/2}
   \exp{\left(-i\lambda\phi_b\sqrt{h}\left(1 -
         \frac{h_0}{2h}\right)\right)} \ .  
 \end{equation}
Hence, the normalisation factor of the density matrix is
divergent. Introducing a cutoff $h < L^2_{\text max}$, and
  integrating $h_0$ from $0$ to $h_c=\lambda^{-2}$, one has
\begin{equation}
  N_b =\int dh_0 \tr(\rho_+^{(h_0,\phi_b)})
  = \int dh dh_0 |\langle h,\phi_b|h_0,0 \rangle_+|^2
\sim \frac{L_{\text max}}{\pi\lambda^3\phi_b} + {\mathcal O}(L_{\text max}^0) \ .
\end{equation}

Clearly, also products of operators will be dominated by semiclassical
intermediate states.
Consider now the trace of the square of the density matrix,
\begin{equation}
  \tr(\rho_+^2) = N_b^{-2} \int dh dh' dh_0 dh'_0
\rho_+^{(h_0,\phi_b)}(h,h') \rho_+^{(h'_0,\phi_b)}(h',h) \ .
\end{equation}
Since the modulus of the semiclassical amplitude \eqref{ampsemi} does
not depend on $h_0$, one obtains
\begin{align}
  \rho_+^{(h_0,\phi_b)}&(h,h') \rho_+^{(h'_0,\phi_b)}(h',h) \nonumber \\
  &\sim \left|\rho_+^{(h_0,\phi_b)}(h,h') \rho_+^{(h'_0,\phi_b)}(h',h)\right| 
   \exp\left(i\frac{\lambda\phi_b}{2\sqrt{h}\sqrt{h'}}
       (h_0-h'_0)(\sqrt{h'}-\sqrt{h})\right) \nonumber \\
&\sim \rho_+^{(h_0,\phi_b)}(h,h) \rho_+^{(h'_0,\phi_b)}(h',h')
   \exp\left(i\frac{\lambda\phi_b}{2\sqrt{h}\sqrt{h'}}
       (h_0-h'_0)(\sqrt{h'}-\sqrt{h})\right) \nonumber \\
&\sim \rho_+^{(h_0,\phi_b)}(h,h) \rho_+^{(h'_0,\phi_b)}(h',h') 
   \left(1 + i\frac{\lambda\phi_b}{2\sqrt{h}\sqrt{h'}}
                   (h_0-h'_0)(\sqrt{h'}-\sqrt{h}))\right. \nonumber \\
& \hspace{4cm}\left.   - \frac{\lambda^2\phi^2_b}{8hh'}
   (h_0-h'_0)^2(\sqrt{h'}-\sqrt{h})^2 + \mathcal{O}((h_0-h'_0)^3)\right) \ .
\end{align}
The integral over the imaginary part vanishes, and one obtains the final result
\begin{align}
\tr(\rho_+^2) &\sim N_b^{-2} \int dh dh' dh_0 dh'_0
\rho_+^{(h_0,\phi_b)}(h,h) \rho_+^{(h'_0,\phi_b)}(h',h') \nonumber\\
   &\hspace{1.5cm} \times\left(1 - \frac{\lambda^2\phi^2_b}{8hh'}
       (h_0-h'_0)^2(\sqrt{h'}-\sqrt{h})^2 + \mathcal{O}((h_0-h_0)^3)\right)
       \nonumber\\
              & < N_b^{-2} \int dh dh_0  \rho_+^{(h_0,\phi_b)}(h,h)
                \int dh' dh'_0 \rho_+^{(h'_0,\phi_b)}(h',h') = 1 \ .
\end{align}
Hence, we find for the semiclassical part of the density matrix
$\tr(\rho_+^2) < 1$, which is the characteristic feature of a mixed
state. It would be interesting to verify this property also for the
full density matrix beyond the semiclassical approximation, and to
understand the role of the cutoff $L_{\text max}$ better. We leave
this for future work.

\section{Prefactor for JT gravity with inflaton}
\label{A:PJTinf}

In section~\ref{sec:JTinf} we discussed the semiclassical wave
function for JT gravity with inflaton. In the following we provide
some details of the derivation.

From the on-shell action \eqref{osGM},
  \begin{equation}\label{osM}
  (\lambda a_0)^{-1}I^{os} = - \left(\phi + c - \kappa\chi
    + \frac{\kappa^2}{2}\right)\da a_0^{-1}  + \frac{\kappa^2}{2} X
  + \frac{1}{2}\left(\chi -\kappa\ln{(a a_0^{-1})}\right)^2 X^{-1} \ , \nonumber
  \end{equation}
and the semiclassical wave function
  \begin{equation}\label{psiaphichi}
  \Psi_0(a,\phi,\chi) =
  C(a,\phi,\chi)\exp{\left(\frac{i}{\hbar}I^{os}(a,\phi,\chi)\right)}
  \ ,
\nonumber
\end{equation}
one obtains the partial differential equation for $z=\ln C$,
\begin{align}
 \Big(& \partial_a 
+ \Big(\Big(\phi + c - \kappa \chi
             + \frac{\kappa^2}{2}\Big) a\da^{-2}
          -\frac{\kappa^2}{2} a_0^2a^{-1}\da^{-2}  \nonumber\\
 &          + \kappa (\chi - \kappa \ln{(a a_0^{-1})})X^{-1}a_0(a\da)^{-1}
 + \frac{1}{2} (\chi - \kappa \ln{(a a_0^{-1})})^2
 X^{-2}a_0^2(a\da)^{-2} \Big)\partial_\phi\nonumber\\
 &+ (\kappa a^{-1} + (\chi  - \kappa \ln{(a a_0^{-1})})
   X^{-1}a_0(a\da)^{-1})\partial_\chi\Big) z(a,\phi,\chi)
   \equiv \mathcal{D} z(a,\phi,\chi) \nonumber\\
        &= - a \da^{-2} - X^{-1}a_0 (a\da)^{-1}\ . \label{Dz}
\end{align}
Here we have divided Eq.~\eqref{pdgd} by $\da$, in order to
parametrise the trajectories directly by $a$.
The corresponding first-order differential equations for the
characteristics read
\begin{align}
 \frac{d\chi}{da} &= \kappa a^{-1} + (\chi  - \kappa \ln{(a a_0^{-1})})
  X^{-1}a_0(a\da)^{-1} \ , \\
  \frac{d\phi}{da} &= \Big(\phi + c - \kappa \chi
             + \frac{\kappa^2}{2}\Big) a\da^{-2}
          -\frac{\kappa^2}{2} a_0^2a^{-1}\da^{-2}  \nonumber\\
 &          + \kappa (\chi - \kappa \ln{(a a_0^{-1})})X^{-1}a_0(a\da)^{-1}
 + \frac{1}{2} (\chi - \kappa \ln{(a a_0^{-1})})^2
   X^{-2}a_0^2(a\da)^{-2} \ ,\\
  \frac{dz}{da} &= - a \da^{-2} - X^{-1}a_0 (a\da)^{-1}\ . \label{zinf}
\end{align}
The solutions for $\chi$ and $\phi$ coincide with the solutions \eqref{inf}
and \eqref{phi} of the equations of motion,
\begin{align}
 \chi &=  \chi_0 X(a) + \kappa \ln{(a a_0^{-1})}
        \ , \label{achi}\\
 \phi &=  \da a_0^{-1}\left(\phi_0-\frac{1}{2}(\kappa^2+\chi_0^2)
   X(a)\right) + \kappa^2\ln{(a a_0^{-1})}
          + \kappa\chi_0 X(a) - c - \frac{1}{2}\chi_0^2 \ . \label{aphi}
\end{align}
They satisfy the boundary conditions $\chi(a_0) = 0$ and $\phi(a_0) = -c
- \chi_0^2/2$, respectively. With $a=a_0\cosh{\lambda t}$, one obtains
the functions
$\chi(a(t)) \equiv \chi(t)$ and $\phi(a(t)) \equiv \phi(t)$, which are
solutions of Eqs.~\eqref{emchi} and \eqref{emphi}, \eqref{hcd},
with the boundary
conditions $\chi(0) = 0$, $\dot{\chi}(0) = \lambda \chi_0$, and 
$\phi(0) = -c-\chi_0^2/2$, $\dot{\phi}(0) = \lambda(\phi_0 +
\kappa\chi_0$), respectively.
For $z(a,z_0)$ one finds
\begin{equation}
  z = z_0 - \ln{(X\da a_0^{-1})} \ .\label{az}
  \end{equation}

Eqs.~\eqref{achi}, \eqref{aphi} and \eqref{az} can be inverted to
obtain the integration constants $\chi_0$ and $\phi_0$ as
functions of $a$, $\chi$ and $\phi$. Since $d\chi_0/da = d\phi_0/da =
0$, $\chi_0(a,\chi)$ and $\phi_0(a,\chi,\phi)$ satisfy the homogeneous part of
Eq.~\eqref{Dz}. $dz_0/da = 0$ implies that $z(a,z_0)$ satisfies the
full inhomogeneous PDE~\eqref{Dz}.
Choosing $z_0= F(\phi_0,\chi_0)$, where $F$ is an arbitrary function,
 yields the solution to the PDE \eqref{pdgd},
  \begin{equation}
    C(a,\phi,\chi) = F(\phi_0  (a,\phi, \chi), \chi_0 (a,\phi,\chi)) a_0\da^{-1} X^{-1} \ .
  \end{equation}

\section{Details of adding a semiclassical inflaton}
\label{A:dPsemiBranching}

When inverting the solutions $\phi(a),\chi(a)$ of the classical equations of motion in terms of the their integration constants and going into the region of large scale factor, one needs to be careful in picking the right branch of arccosine function appearing in the inflaton solution. We see this by looking at Eqs.~\eqref{inf} and \eqref{phi}. Using these, one obtains for the integration
constants $\chi_0$ and $\phi_0$ at large $a$,

\begin{align}
  X(a) &= \pm \frac{\pi}{2}\mp\mathcal{O}(a^{-1}) \ , \\
  \chi_0 (a,\chi) &= X^{-1} \left(\chi - \kappa\ln{(a}{a^{-1}_0)}
  \right)  \ , \\
  \phi_0 (a, \phi, \chi) &= \frac{\phi a_0}{\da}
   \pm \frac{\pi\kappa^2}{4}  \pm\frac{1}{\pi}(\chi - \kappa\ln{(a a_0^{-1})})^2 + \mathcal{O}(a^{-1}) \label{phi0infDoubleBranch}\ .
\end{align}
The subtlety shows itself in the two signs above. They pertain to the ambiguity arising from inverting the cosine function: Within the half-period between $-\pi/2$ and $\pi/2$ where the cosine is positive semi-definite, inverting it near cosine-value zero gives two possible regimes: For $x\to0^+$ the function $\arccos(x)$ is either $\pi/2-{\cal O}(x)$ or $-\pi/2+{\cal O}(x)$. To describe an initial condition which has the inflaton $\chi$ growing steadily with $a$ increasing above its initial value $a_0\,$, one needs to choose the sign of $\chi_0$ correlated with the choice of the branch for the inversion of the cosine. We will now fix the choice of sign by a physical argument concerning the structure of the probability distribution constructed using the measure $dP$.

Using this measure, we obtain from
Eqs.~\eqref{Cmtyinf} and \eqref{phi0inf} for the
probability distribution, 
up to terms of relative order $\mathcal{O}(a^{-1})$,
\begin{align}
  dP(a,\chi;\phi_b) &=|C|^2 dh \nonumber\\
  &= C_0^2\phi_0 \frac{a_0^2}{\Delta_a^2}X^{-2}\nonumber\\
&\simeq \frac{4C_0^2}{\pi^2}  \left(\frac{\phi_b a_0}{\da}
\pm \frac{\pi\kappa^2}{4}  \pm
\frac{\kappa^2}{\pi}\left(\frac{\chi}{\kappa}
  - \ln{(a a^{-1}_0)}\right)^2\right) \frac{a^2_0}{\da^2}ada d\chi \ .
\end{align}
A semiclassical probability distribution should display a local maximum on-shell, that is, along the trajectory carved out by a solution to the classical equations of motion for a given set of initial conditions, which here is the solution for $a=a(\chi)$ in the large-$a$ limit given by Eq.~\eqref{dclass}. The above expression conforms to this general rule if we choose $\arccos(x)=-\pi/2+{\cal O}(x)$  for $x\to0^+$. Hence, the probability distribution becomes the expression~\eqref{dPinfSC} in the main text.

\bibliographystyle{utphys}
\bibliography{JT}
%JHEP.bst
\end{document}